\documentclass{amsart}
\usepackage{rotating} 
\usepackage[americanvoltages]{circuitikz}
\usepackage{siunitx}
\usepackage{amsmath, amssymb, amsfonts }
\usepackage{subfig}
\usepackage{array, makecell}
\usepackage{adjustbox}
\usepackage{multirow}
\usepackage[margin=1.3in]{geometry}
\usepackage[foot]{amsaddr}
\usepackage{algorithmic}
 \usepackage{graphicx}
 \usepackage{textcomp}
 \usepackage{multirow}
 \usepackage{cancel}
 
 \newcommand{\junk}[1] {}
 
 \newcommand{\D}{D^{\nu}(s)}
 \newcommand{\N}{N^{\nu}(s)}
 \newcommand{\Dext}{d_0^{\nu}+\sum^n_{i=1}\dfrac{d_i^\nu}{s-a^\nu_i}}

 \newcommand{\Dextnonu}{d_0+\sum^n_{i=1}\dfrac{d_i}{s-a_i}}
 \newcommand{\Dextznonu}{\frac{d_0}{s}+\sum^n_{i=1}\dfrac{d_i}{s\cdot(s-a_i)}}
 \newcommand{\Nextnonu}{c_0+\sum^n_{i=1}\dfrac{c_i}{s-a_i}}
 
 \newcommand{\heaviside}{\theta}

\begin{document}
\title[Time-Domain Vector Fitting for Boundary Conditions Estimation ]{A Vector Fitting Approach for the Automated Estimation of Lumped Boundary Conditions of 1D
	Circulation Models}
\author[Elisa Fevola et al.]{Elisa~Fevola$^1$, Tommaso Bradde$^{1}$, Piero Triverio$^2$ and Stefano~Grivet-Talocia$^1$}
\address{$^1$ Department of Electronics and Telecommunications, Politecnico di Torino, Torino, Italy}
\address{$^2$ Department of Electrical \& Computer Engineering, Institute of Biomedical Engineering, University of Toronto, Toronto, Canada}
\begin{abstract}
The choice of appropriate boundary conditions is a crucial step in the development of cardiovascular models for blood flow simulations.
The three-element Windkessel model is usually employed as a lumped boundary condition, providing a reduced order representation of the peripheral circulation. However, the systematic estimation of the Windkessel parameters remains an open problem. Moreover, the Windkessel model is not always adequate to model blood flow dynamics, which often require more elaborate boundary conditions. In this study, we propose a method for the estimation of high order boundary conditions, including the Windkessel model, and we investigate their use. 
The proposed technique is based on Time-Domain Vector Fitting, a modeling algorithm that, given samples of the input and output of a system, such as pressure and flow waveforms, can derive a differential equation approximating their relation. 
The capability of the proposed method is tested on a 1D circulation model consisting of the 55 largest arteries, to demonstrate its accuracy and the usefulness of estimating boundary conditions with order higher than the traditional Windkessel models. The proposed method is verified against other common estimation techniques, and its robustness in parameter estimation is verified in presence of noisy data and of physiological changes of aortic flow rate induced by mental stress.
Results suggest that the proposed method is able to accurately estimate boundary conditions of arbitrary order. 
Higher order boundary conditions can improve the accuracy of cardiovascular simulations, and Time-Domain Vector Fitting can automatically estimate them.

\end{abstract}

\maketitle
\section{Introduction} \label{sec:intro}
{C}{omputational} models of the cardiovascular system have become a valuable tool for the study and investigation of cardiovascular diseases~\cite{formaggia2010cardiovascular}.
Since a simulation of the entire cardiovascular system is computationally expensive, cardiovascular models usually include only a specific region of interest. The excluded regions are taken into account by choosing appropriate boundary conditions (BCs), which must provide a realistic representation of the haemodynamics in the rest of the circulatory system.  Boundary conditions have been shown to largely affect flow rates, pressure distribution and important haemodynamic indicators, such as wall shear stress~\cite{pirola2017choice, morbiducci2013inflow}. For this reason, the selection of proper inlet and outlet boundary conditions  that can realistically reproduce blood flow dynamics is particularly important. 

Different types of outlet boundary conditions have been proposed~\cite{shi2011review}. Among these, the most commonly adopted in order of complexity are:
\begin{itemize}
	\item boundary conditions that simply prescribe a specific value for pressure or flow rate at the outlets~\cite{parker1990forward};
	\item constant resistances, which result in a linear algebraic relation between pressure and flow rate~\cite{rooz1982finite, wang2004wave, wan2002one};
	\item boundary conditions that impose a differential relation between pressure and flow rate, usually represented as equivalent lumped parameter networks. The latter can be classified according to their order, which corresponds to the number of storage elements (capacitors and inductors) present in the circuit, or equivalently to the order of the corresponding differential equation. A typical example is the three-element Windkessel model (3WK)~\cite{segers2008three}, a circuit of order one displayed in Fig.~\ref{fig:3wk}. Higher order Windkessel models, containing additional capacitors and inductors, have also been proposed~\cite{jonavsova2021relevance}.
\end{itemize}

The choice of the best model for outlet boundary conditions is generally the result of a trade-off between accuracy, model complexity, and number of parameters to estimate. For this reason, the most popular choice is the three-element Windkessel model, also known as RCR model (see Fig.~\ref{fig:3wk} and Sec.~\ref{subsec:wk_method}). \junk{, where capacitance $C$ mimics the elastic properties of arteries, while the two resistance elements $R_1$ and $R_2$ represent the vascular resistance, and $P_d$ represents the distal pressure.} Even if the number of parameters in the Windkessel is limited, obtaining an accurate estimate is not straightforward.
A simple, yet expensive approach consists in identifying reasonable ranges for each parameter, and then refining the choice by means of an iterative tuning procedure~\cite{westerhof2009arterial}.
If both pressure and flow data are available, more advanced and systematic approaches are generally used, where Windkessel models are fitted to available data by means, for example, of the simplex search method~\cite{segers2008three}, or by least-square minimization~\cite{romarowski2018patient}.
Similarly, in~\cite{alastruey2008lumped}, the terminal Windkessel resistances are estimated from mean pressure and outflow measurements at each terminal vessel, while terminal compliances are obtained by distributing the total peripheral compliance according to the cross-sectional areas of the outlets. The method proposed in~\cite{epstein2015reducing}, instead, selects parameters of the Windkessel models such that the net resistance and total compliance of the entire system are preserved.
Other solutions resort to a non-iterative subspace model identification algorithm~\cite{kind2010estimation}, or to other data-assimilation techniques, such as Kalman filtering~\cite{vignon2014methodological} and optimal control~\cite{fevola2021optimal}, but their applicability is limited by their high computational cost.
Overall, the existing solutions for the estimation of Windkessel parameters tend to be either empirical, or time consuming. Moreover, most of the available approaches are suitable only for the estimation of first order boundary conditions, such as the 3WK model, and are hard to generalize to higher order. Higher order BCs, in fact, have been proven to be more accurate and realistic than the three-element Windkessel model~\cite{segers2008three, stergiopulos1999total}, but the difficulty in estimating a larger number of parameters has limited their diffusion.

In this paper, we propose a novel approach for the automated estimation of boundary conditions of arbitrary order. The proposed method is based on the Time-Domain Vector Fitting algorithm (TDVF), which approximates the behavior of a system by means of differential equations relating input and output~\cite{grivet2003package, bradde2021handling}. Supposing that pressure and flow rate samples are available at the truncation location, where the boundary condition must be imposed, the Time-Domain Vector Fitting can provide a boundary condition of arbitrary order relating pressure and flow rate very accurately.
In the case of a model of order one, the proposed method provides an automated way to estimate the Windkessel parameters. For orders higher than one, instead, the model is represented as a differential relation between pressure and flow rate, which can be used as a boundary condition to Navier-Stokes equations, and is easy to implement in computational fluid dynamics (CFD) solvers.

We assess the capability of the proposed method on a 1D circulation model consisting of the 55 largest systemic arteries~\cite{alastruey2012arterial}, by truncating some portions of the system and replacing them with boundary conditions of increasing order estimated with Vector Fitting (VF). Experimental results show that boundary conditions estimated with the proposed algorithm provide accurate pressure and flow rates at the truncation locations, making VF a promising candidate for parameter estimation in cardiovascular models. For Windkessel models, VF is compared to two other methods in the literature, one preserving the net resistance and total compliance of the original 55-artery system~\cite{epstein2015reducing}, and the other based on the Nelder-Mead simplex algorithm~\cite{lagarias1998convergence}. Overall,  VF produces comparable, or better, results. 
The main advantage of the proposed approach is that it can easily estimate conditions of order higher than one, and results will show that these can provide increased accuracy. Lastly, we verify that the proposed technique is able to accurately fit pressure and flow waveforms affected by noise down to 20 dB of signal-to-noise ratio, and that the estimated BCs remain valid in presence of physiological changes of the input waveforms (e.g., in case of mental stress~\cite{charlton2018assessing, celka2020influence}).

\section{Methodology} \label{sec:methodology}
\begin{figure}
	\includegraphics[width=0.95\columnwidth,clip]{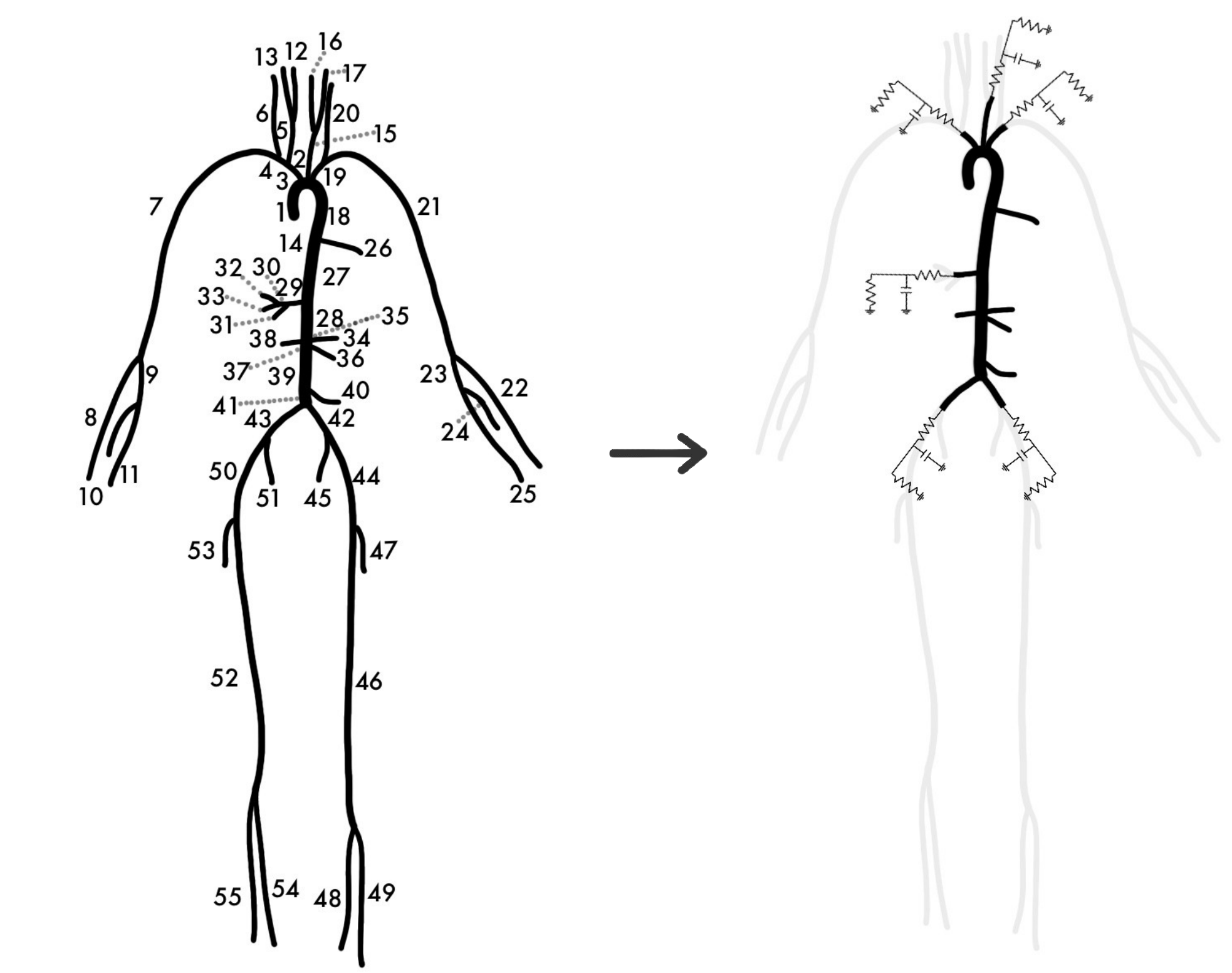}
	\caption{Left: schematic representation of the complete 55-artery network. Right: representation of the reduced model after boundary conditions estimation with Vector Fitting. The arterial segments are reduced from 55 to 21, and the truncated parts of the system (in grey) are substituted with the estimated boundary conditions. These could be Windkessel models, as depicted here, or models of higher order.}
	\label{fig:55_artery_network}
\end{figure}

In this section, the Time-Domain Vector Fitting algorithm will be introduced, together with the proposed formulation for boundary conditions estimation. The goal of this procedure is represented in Fig.~\ref{fig:55_artery_network}, where we want to move from a model representing the systemic arterial system (Fig.~\ref{fig:55_artery_network}, left), to a reduced version where part of the vasculature has been removed and substituted by properly estimated boundary conditions (Fig.~\ref{fig:55_artery_network}, right). The latter could be Windkessel models, as displayed in~Fig.~\ref{fig:55_artery_network}, or general boundary conditions of higher order.

In the following subsections, first the three-element Windkessel model will be briefly reviewed (Sec.~\ref{subsec:wk_method}), and then generalized to a form suitable for the VF algorithm (Sec.~\ref{subsec:general_wk}). The VF algorithm for the estimation of boundary conditions of arbitrary order will be presented in Sec.~\ref{subsec:vf_wk}, and the implementation of the obtained boundary conditions in CFD solvers will be presented in Sec.~\ref{subsec:bc_implementation}.

\subsection{The three-element Windkessel model}\label{subsec:wk_method}
The three-element Windkessel model was first introduced by Westerhof et al.~\cite{westerhof1969analog}. Its circuit interpretation includes three elements, as displayed in Fig.~\ref{fig:3wk}: the capacitor $C$ models the storage properties of arteries, the resistor $R_1$ represents the proximal resistance of the arterial network, while the resistor $R_2$ models the resistance of the distal circulation. Moreover, a distal pressure contribution $P_d$ is also included, in order to represent the pressure at which flow to the microcirculation ceases~\cite{alastruey2012arterial}.

\begin{figure}[t]
	\centering
	\begin{circuitikz} [american]
		\ctikzset{voltage/distance from node=.2}
		\ctikzset{voltage/distance from line=.02}
		\ctikzset{voltage/bump b/.initial=.0}
		\draw
		(0,0) node[anchor=east]{}
		to[short, o-] (3,0)
		to[C, l=$C$] (3,2)
		to[R, l_=$R_{1}$, i_<=$q(t)$, -o] (0,2)  
		(3,2) to[R, l=$R_{2}$] (5.5, 2)
		to[battery2, v=$P_d$] (5.5, 0) node[ground] {}
		(5.5, 0) -- (3, 0)
		
		(0,0) to [open,v<=$p(t)$] (0,2)
		;\end{circuitikz}
	\caption{The three-element Windkessel model used as outlet boundary condition. The circuit includes the proximal resistance $R_1$, the distal resistance $R_2$, the capacitance $C$ and the distal pressure $P_d$.} \label{fig:3wk}
	
	\centering
	\begin{circuitikz} 
		\ctikzset{voltage/distance from node=.2}
		\ctikzset{voltage/distance from line=.02}
		\ctikzset{voltage/bump b/.initial=.0}
		\draw
		(0,0) node[anchor=east]{}
		to[short] (3,0)
		to[C, l=$C$] (3,2)
		to[R, l_=$R_{1}$] (0,2)  
		(3,2) to[R, l=$R_{2}$] (5.5, 2)
		(-1,2) to[V, v<=$\widetilde{p}_d(t)$] (0, 2) 
		(-2,2) to[short, o-, i=$q(t)$] (-1, 2) 
		(5.5,0)  to[short] (5.5,2)
		(5.5, 0) -- (2, 0)
		(-2,0) to[short, o-] (0,0)
		(-2,0) to [open, v<=$p(t)$](-2,2)
		(5.5,0)  node[ground]{}
		;\end{circuitikz}   
	\caption{Equivalent Windkessel model, obtained by relocating the distal pressure contribution.}
	\label{fig:windkesselModifiedIO}

	\centering
	\begin{circuitikz} 
		\ctikzset{voltage/distance from node=.2}
		\ctikzset{voltage/distance from line=.02}
		\ctikzset{voltage/bump b/.initial=.0}
		\draw
		(0,0) node[anchor=east]{}
		to[short] (3,0)
		to[C, l=$C$] (3,2)
		to[R, l_=$R_{1}$] (0,2)  
		(3,2) to[R, l=$R_{2}$] (5.5, 2)
		(-1,2) to[battery2, v=$P_d$] (0, 2) 
		(-2,2) to[short, o-, i=$q(t)$] (-1, 2) 
		(5.5,0)  to[short] (5.5,2)
		(5.5, 0) -- (2, 0)
		(-2,0) to[short, o-] (0,0)
		(-2,0) to [open, v<=$p(t)$](-2,2)
		(5.5,0)  node[ground]{}
		;\end{circuitikz}   
	\caption{Approximate Windkessel model.}
	\label{fig:windkesselApprox}

	\centering
	\begin{circuitikz}[american voltages] \draw (0,0)
		node[draw,minimum width=2cm,minimum height=2.6666cm] (load) {$H(s)$}
		($(load.west)!0.75!(load.north west)$) coordinate (la)
		($(load.west)!0.75!(load.south west)$) coordinate (lb)
		(lb) to[short] ++(-0.5,0) coordinate (b) node[below] {}
		to[short,-o] ++(-3,0) coordinate (VThb)
		to[open,v<=$p(t)$] (VThb |- la)
		to[short,o-,i=$q(t)$] ++(1,0)
		to[battery2=$P_d$] ++(1.5,0) coordinate (VTht)
		to[short] (la);
	\end{circuitikz}
	\caption{Proposed high-order boundary condition model.}
	\label{fig:high-order}
\end{figure}

The Windkessel model relates pressure $p(t)$ to flow rate $q(t)$ by means of the differential equation
\begin{equation}\label{eq:windkesselDifferential}
q(t)\left(1+ \frac{R_1}{R_2}\right) + C R_1 \frac{d q}{d t} = \frac{p(t) - P_d}{R_2} + C\frac{d p}{d t},
\end{equation}
whose derivation from the equivalent circuit of Fig.~\ref{fig:3wk} is straightforward by exploiting the equivalence between fluid dynamics quantities (pressure, flow rate) and electrical quantities (potential, current). Estimating the Windkessel parameters consists in determining the optimal values for $R_1$, $R_2$, $C$ and $P_d$ in~\eqref{eq:windkesselDifferential} that best approximate the time domain evolution of the pressure and flow rate at the outlet.

\subsubsection{Laplace-domain formulation}

The following derivations and generalizations are best described in the Laplace domain~\cite{widder2015laplace}. The Laplace transform $\mathcal{L}$ is a standard mathematical tool that converts linear differential equations into algebraic equations, leading to a drastic simplification in both solution and interpretation of differential models.

Let us denote with $s$ the Laplace variable (representing the time derivative operator $d/dt$), and define the Laplace transforms of $p(t)$ and $q(t)$ as $P(s)$ and $Q(s)$, respectively. Assuming vanishing initial conditions at $t=0$, the Laplace transform of~\eqref{eq:windkesselDifferential} is
\begin{equation}\label{eq:differentialToLaplace}
Q(s)\left(1+ \frac{R_1}{R_2}\right) + s C R_1 Q(s) = \frac{P(s)}{R_2} -\frac{ P_d}{sR_2} + sCP(s),
\end{equation}
which is an algebraic relation between pressure and flow rate, parameterized by the constants $R_1$, $R_2$, $C$, and $P_d$. The distal pressure $P_d$ can be interpreted both as a free parameter, but also as an extra (constant) input, with specific reference to the circuit interpretation of Fig.~\ref{fig:3wk} where it is represented as a voltage source.

Equation~\eqref{eq:differentialToLaplace} can be rewritten as
\begin{equation}\label{eq:windkesselFreq_1}
P(s) =  H(s)Q(s)+H_d(s)\frac{P_d}{s},
\end{equation}
where $H(s)$ and $H_d(s)$ are the two transfer functions
\begin{equation}
H(s)= R_1 +  \frac{R_2}{sR_2C+1}, \quad 
H_d(s)=  \frac{1}{sR_2C+1} 
\end{equation}
These are two first-order rational functions of the Laplace variable $s$, whose \emph{order} is defined as the degree of the denominator. This is coherent with the differential equation~\eqref{eq:windkesselDifferential}, which includes only first-order derivatives. To enable the generalization proposed in this paper, we rewrite these transfer functions in the general pole-residue (partial fraction) form as
\begin{equation}\label{eq:structureTargetTf1PR} 
H(s)=c_0 + \frac{c_1}{s-a} \quad
H_d(s)=  \frac{b_1}{s-a} ,
\end{equation}
where the pole $a$, the residues $c_1$, $b_1$, and the direct coupling constant $c_0$ can be uniquely related to the Windkessel parameters through
\begin{align}\label{eq:poleResiduesToParameters}
R_1 = c_0,\quad 
R_2=-\frac{c_1}{a}, \quad
C=\frac{1}{c_1},\quad {\rm with} \quad b_1=-a.
\end{align}

\subsection{Generalization to high order boundary conditions}\label{subsec:general_wk}

In this section we show how~\eqref{eq:windkesselFreq_1} can be generalized to arbitrary order, in a way that will facilitate the estimation of its coefficients using the VF algorithm, presented in Sec.~\ref{subsec:vf_wk}.
The proposed high order boundary conditions require a number of steps to be properly defined, starting from the standard 3WK model. These steps are discussed in the three following sections. In particular, the first two steps eliminate the requirement of estimating two transfer functions, by modifying the structure of the boundary condition model to a single transfer function $H(s)$. The latter is defined in terms of a high-order transfer function in pole-residue form in the third step. We will see that this model structure simplifies estimation of the parameters in Sec.~\ref{subsec:vf_wk}.

\subsubsection{Relocation of the distal pressure contribution}\label{sec:wk_output}

A well known result in circuit theory states that any linear and time invariant circuit with one port and internal sources can be transformed into an equivalent circuit, consisting of the series of an impedance and a voltage source. This result, known as Thevenin theorem~\cite{chua1987linear}, applies also to the present application case. When applied to the 3WK circuit of Fig.~\ref{fig:3wk}, we obtain the circuit of Fig.~\ref{fig:windkesselModifiedIO}, where the equivalent source term is denoted as $\widetilde{p}_d(t)$. A full equivalence with the 3WK model is established in the Laplace domain by setting
\begin{equation}
\widetilde{P}_d(s) = H_d(s)\frac{P_d}{s}
\end{equation}
so that~\eqref{eq:windkesselFreq_1} can be restated as
\begin{equation}\label{eq:windkesselFreq}
P(s) =  H(s)Q(s)+\widetilde{P}_d(s).
\end{equation}
With these definitions, the two circuits in Fig.~\ref{fig:3wk} and Fig.~\ref{fig:windkesselModifiedIO} are indistinguishable in terms of the induced relationship between $P(s)$ and $Q(s)$.

Since $P_d$ is constant, applying the inverse Laplace transform to $\widetilde{P}_d(s)$ leads to
\begin{equation}\label{eq:pd_tilde}
\widetilde{p}_d(t) = P_d \, \left( 1- e^{at} \right) \theta(t)
\end{equation}
where we used~\eqref{eq:structureTargetTf1PR}-\eqref{eq:poleResiduesToParameters}, and where $\theta(t)$ is the unit step (Heaviside) function. The signal $\widetilde{p}_d(t)$ converges exponentially to the asymptotic value $P_d$ with an initial transient, whose duration is related to the time constant $\tau = -1/a = R_2 C$.

\subsubsection{Approximation of early-time transient behavior}\label{sec:wk_approx}

The proposed generalization to higher order requires an approximation, which is motivated and discussed below. Time-domain cardiovascular simulations are usually initialized to a vanishing initial state for all variables (pressure and flow rate in our case). However, the solution of practical and clinical interest is the periodic state operation that arises due to pulsating input excitation, which is usually applied in form of a predefined flow rate at the inlet. Such periodic state is reached after an initial transient, which is inevitably required by the numerical solvers, and which is generally disregarded when interpreting the results of the simulation.

\begin{figure}
	\centering
	\includegraphics[width=\columnwidth]{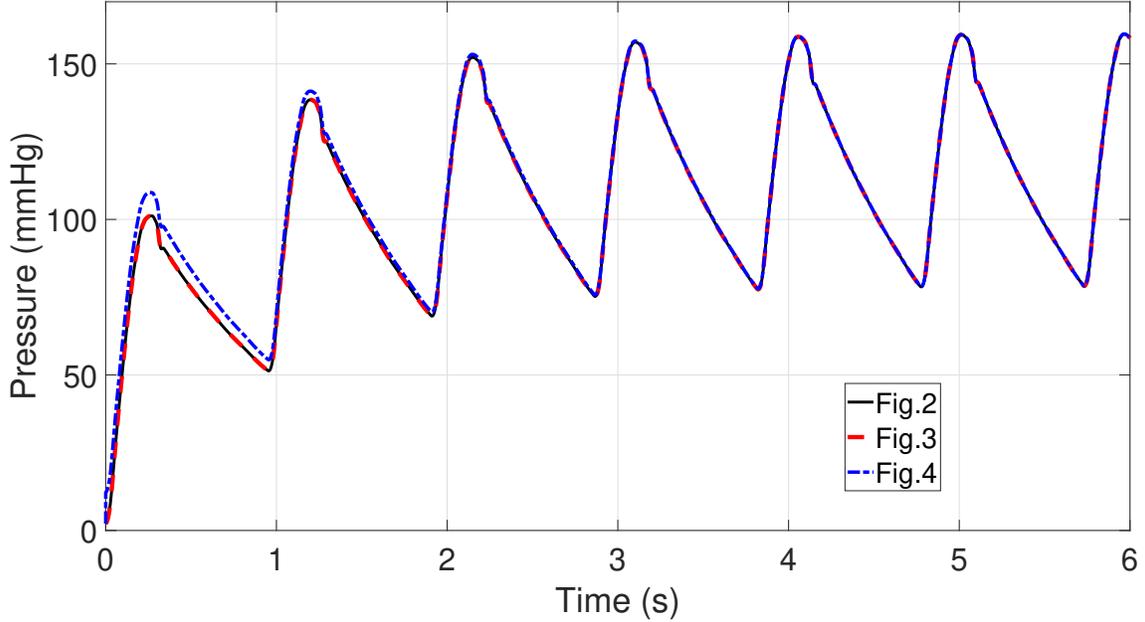}
	\caption{Pressure signals obtained by exciting the three boundary condition models in Fig.~\ref{fig:3wk} (black line), Fig.~\ref{fig:windkesselModifiedIO} (red dashed line) and Fig.~\ref{fig:windkesselApprox} (blue line) with the same inlet flow excitation signal.}
	\label{fig:wk_gen_check}
\end{figure}

Given the above observation, and noting that the equivalent source $\widetilde{p}_d(t)$ differs from its asymptotic value $P_d$ only during the initial transient, we replace $\widetilde{p}_d(t)$ with $P_d$ in the circuit of Fig.~\ref{fig:windkesselModifiedIO}, obtaining the approximate Windkessel model depicted in Fig.~\ref{fig:windkesselApprox}. This operation corresponds to redefining $H_d(s)=1$ in~\eqref{eq:windkesselFreq}, which is then approximated as
\begin{equation}\label{eq:windkesselFreqApprox}
P(s) \approx  H(s)Q(s)+\frac{P_d}{s}.
\end{equation}
A strict equivalence with the initial 3WK model of Fig.~\ref{fig:3wk} no longer holds, but the only difference between the two formulations occurs at early times. When the initial transient is extinguished, the periodic states obtained with the two models are identical. This is confirmed by Fig.~\ref{fig:wk_gen_check}, where the pressure signals obtained by exciting the three models in Fig.~\ref{fig:3wk}, Fig.~\ref{fig:windkesselModifiedIO} and Fig.~\ref{fig:windkesselApprox} with the same inlet flow excitation are depicted. The first two responses are identical in light of the full equivalence of the corresponding models. The response of the approximate model (blue line) asymptotically converges to the other two signals after the initial transient is extinguished. We conclude that, if only the periodic state operation is required, all discussed boundary condition models are equivalent.

\subsubsection{Generalization to arbitrary order}\label{sec:w_high_order}

Assuming the approximation discussed in Sec.~\ref{sec:wk_approx}, generalization to higher order boundary conditions becomes straightforward. We simply redefine the transfer function $H(s)$ in~\eqref{eq:windkesselFreqApprox} as a higher order rational function, expressed in pole-residue form as
\begin{equation}\label{eq:H_high_order}
H(s)=c_0 + \sum_{i=1}^n\frac{c_i}{s-a_i}.
\end{equation}
\junk{Vector Fitting provides a way to estimate the coefficients of~\eqref{eq:H_high_order} from pressure and flow rate measurements in an accurate and robust way. Based on~\eqref{eq:H_high_order}, it is also possible to represent~\eqref{eq:windkesselFreqApprox} as an equivalent electrical circuit using techniques applied from the literature~\cite{grivet2015passive,antonini2003spice}. However, if the goal is to implement these boundary conditions in a 1D or 3D CFD simulator, this may not be necessary, since~\eqref{eq:H_high_order} can be easily converted to a differential equation.}
Based on~\eqref{eq:H_high_order}, the representation~\eqref{eq:windkesselFreqApprox} is easily converted to a set of coupled differential equations for direct inclusion as boundary conditions in 1D or 3D CFD solvers, see later Sec.~\ref{subsec:bc_implementation}. Therefore, the proposed high-order boundary condition model should be regarded as a black-box representation of the differential relation between outlet pressure and flow variables, characterized by richer dynamics and generally allowing for more accurate numerical results. These claims will be demonstrated by the numerical examples of Sec.~\ref{sec:results}. \junk{We remark that inclusion of~\eqref{eq:H_high_order} in CFD solvers is straightforward, as discussed later in Sec.~\ref{subsec:bc_implementation}.}

We close this section by providing an interpretation for the presence of $s$ at the denominator associated to $P_d$ in~\eqref{eq:windkesselFreqApprox}. Since the Laplace transform of the unit step $\theta(t)$ is $\mathcal{L}\{\theta(t)\} = 1/s$, we see that the distal pressure term in~\eqref{eq:windkesselFreqApprox} can be interpreted as a time-domain source $\widetilde{p}_d(t)=P_d\, \theta(t)$. Therefore, we see that the proposed high-order model assumes that the distal pressure contribution $P_d$ is applied instantaneously at $t=0$, rather than through an exponential transient~\eqref{eq:pd_tilde}. As discussed above, this difference is irrelevant when considering only the periodic state solution.

\subsection{Time-Domain Vector Fitting for boundary conditions estimation}\label{subsec:vf_wk} 

We now discuss how the parameters of the proposed high-order boundary conditions~\eqref{eq:windkesselFreqApprox}-\eqref{eq:H_high_order} can be automatically estimated from time series of pressure $p(t)$ and flow rate $q(t)$ at some vessel outlet. We assume that samples of these signals are available as
\begin{equation}\label{eq:data}
p(t_k),\;q(t_k), \quad k=0,\dots, K, \quad t_0=0,
\end{equation}
with a constant sampling rate $\Delta t = t_{k+1}-t_k$ and vanishing initial conditions $p(t_0)=q(t_0)=0$. Generalization to non-vanishing initial conditions will be provided in Sec.~\ref{subsubsec:in_vivo}.

\subsubsection{Model parameterization}

The proposed approach is based on the following model structure
\begin{align}\label{eq:modelstructure}
&H(s)=\frac{N(s)}{D(s)},\\
&N(s)=c_0+\sum^n_{i=1}\dfrac{c_i}{s-a_i} \label{eq:numDef}, \\
&D(s)=d_0+\sum^n_{i=1}\dfrac{d_i}{s-a_i} \label{eq:denDef}.
\end{align}
The transfer function $H(s)$ is expressed as a ratio of two rational functions sharing the same set of common poles $\{a_i\}$ with unknown residues $\{c_i\}$ and $\{d_i\}$. A simple algebraic simplification shows that these poles eventually cancel out: the poles $\{a_i\}$ are simply instrumental variables on which we construct the identification algorithm. The expressions~\eqref{eq:modelstructure}-\eqref{eq:denDef} provide a parameterization of all proper rational functions with order $n$.

\subsubsection{The VF iteration}

Let us start assuming that the poles $\{a_i\}$ in~\eqref{eq:numDef} and~\eqref{eq:denDef} are known, so that $H(s)$ is parameterized only by the residues $\{c_i\}$, $\{d_i\}$ of numerator and denominator, respectively. These unknowns are computed by enforcing~\eqref{eq:windkesselFreqApprox} as a fitting condition, based on the available pressure and flow samples.
Using~\eqref{eq:modelstructure}, we rewrite~\eqref{eq:windkesselFreqApprox} as
\begin{equation}\label{eq:linearizedFitting}
D(s) P(s) \approx N(s) Q(s) +  \frac{P_d}{s}D(s),
\end{equation}
which is obtained by multiplying both sides by the (unknown) denominator $D(s)$. Plugging~\eqref{eq:numDef} and~\eqref{eq:denDef} into~\eqref{eq:linearizedFitting} leads to the relation
\begin{multline}\label{eq:linExpanded}
\left( \Dextnonu \right)P(s) \approx \left( \Nextnonu \right) Q(s) +\\ +P_d\left( \Dextznonu \right).
\end{multline}
The above can be expressed in time domain by applying the inverse Laplace transform to both sides, obtaining 
\begin{align}\label{eq:TDlinearization}
d_0 \cdot p (t) + \sum^n_{i=1} d_i \cdot {p}_i (t)&
\approx   c_0 \cdot q(t) + \sum^n_{i=1}   c_i \cdot  {q}_i (t)  +        \\
& + P_d \,d_0 \cdot \heaviside(t) + \sum^n_{i=1} P_d \, d_i \cdot {\heaviside}_i(t),   \nonumber
\end{align}
where we used the shorthand notation
\begin{equation}\label{eq:filtered_z}
{z}_i(t) = \int_{0}^t e^{a_i(t-\tau)} z(\tau) d \tau
\end{equation}
for any signal $z(t)$. In order to  render the approximation problem linear in the decision variables, we introduce the new set of dummy variables $b_i = P_d\, d_i$ and writing~\eqref{eq:TDlinearization} for all discrete time samples $t=t_k$ leads to a homogeneous linear least squares problem in the unknowns $\{c_i\}$, $\{d_i\}$, $\{b_i\}$, which takes the compact form
\begin{equation}\label{eq:VFleastsquares}
A\,x\approx 0, \quad A=\begin{bmatrix}
-\Phi & \Gamma & \Theta 
\end{bmatrix},\quad x=\begin{bmatrix}
d \\ c \\ b
\end{bmatrix}
\end{equation}
where 
\begin{align}
\Phi &=\begin{bmatrix}
p(t_0)&{p}_1(t_0)&\hdots&{p}_n(t_0)\\
\vdots&\vdots&\ddots&\vdots\\
p(t_K)&{p}_1(t_K)&\hdots&{p}_n(t_K)
\end{bmatrix}, & d=\begin{bmatrix} d_0 \\d_1\\ \vdots\\ d_n \end{bmatrix}\\
\Gamma&=\begin{bmatrix}
q(t_0)&{q}_1(t_0)&\hdots&{q}_n(t_0)\\
\vdots&\vdots&\ddots&\vdots\\
q(t_K)&{q}_1(t_K)&\hdots&{q}_n(t_K)
\end{bmatrix} , & c=\begin{bmatrix} c_0 \\c_1\\ \vdots \\ c_n \end{bmatrix}\\
\Theta&=\begin{bmatrix}
\heaviside(t_0)&{\heaviside}_1(t_0)&\hdots&{\heaviside}_n(t_0)\\
\vdots&\vdots&\ddots&\vdots\\
\heaviside(t_K)&{\heaviside}_1(t_K)&\hdots&{\heaviside}_n(t_K)
\end{bmatrix}, & b=\begin{bmatrix} b_0 \\b_1\\ \vdots \\ b_n \end{bmatrix}.
\end{align}

The solution of~\eqref{eq:VFleastsquares} is computed by enforcing $x\neq 0$ so that the trivial all-zero solution is avoided. This can be done e.g. by computing the Singular Value Decomposition (SVD)~\cite{trefethen1997numerical} of the matrix A,
\begin{equation}
A=U\Sigma V^T
\end{equation}
and choosing $x$ as the last column of $V$, i.e., the right singular vector associated with the least singular value. Alternatively, a non-triviality constraint can be introduced in the problem, as in~\cite{gustavsen2006improving}.

\subsubsection{Pole relocation}
The above procedure determines the optimal set of coefficients $\{c_i\}$, $\{d_i\}$ given a prescribed set of numerator and denominator poles $\{a_i\}$, considered as known quantities. We now consider also these poles as unknowns to be determined. We will see below that the $\{a_i\}$ play the role of estimates for the poles of $H(s)$, which are iteratively refined through a process denoted as \emph{pole relocation}~\cite{gustavsen1999rational,grivet2015passive}.

An iteration with index $\nu$ is set up. At the first iteration $\nu=0$, the starting poles $\{a_i^{0}\}$ are initialized with a set of randomly distributed values throughout the expected frequency band of the model~\cite{gustavsen1999rational}. At any given iteration $\nu$, the set of current poles $\{a_i^{\nu}\}$ is used to construct and solve the least squares system~\eqref{eq:VFleastsquares}. Let us denote as
\begin{equation}\label{eq:Dnext}
\D = \Dext
\end{equation}
the model denominator defined by the coefficients $\{d_i^\nu\}$ resulting from the least squares solution. Since the zeros of $D(s)$ provide the poles of $H(s)$, we define the poles for the next iteration as the zeros of $\D$
\begin{equation}\label{eq:newPoles}
\{a^{\nu+1}\}=\{a_i: D^{\nu}(a_i)=0\}.
\end{equation}
Evaluation of these zeros amounts to solving a small eigenvalue problem, see~\cite{gustavsen1999rational,grivet2015passive}.

In summary, the proposed algorithm involves solving~\eqref{eq:VFleastsquares} and redefining poles through~\eqref{eq:newPoles} for $\nu=0,1,\dots$, until the set $\{a_i^{\nu}\}$ stabilizes. Under this convergence condition, poles and zeros of $\D$ coincide so that $\D=d_0^{\nu}$,  and the model~\eqref{eq:modelstructure} reduces to the numerator $\N$, characterized by poles $\{a_i^{\nu}\}$ and residues $\{c_i^{\nu}\}$. This algorithm can be regarded as an extension of the well-known TDVF scheme~\cite{grivet2003package}, suitably modified to account for the presence of the (unknown) distal pressure term, which produces the matrix block $\Psi$ and the additional dummy unknowns $\{b_i\}$ in~\eqref{eq:VFleastsquares}.

\subsubsection{Estimation of the distal pressure}

Once $H(s)$ is available from the above pole relocation iteration, the distal pressure $P_d$ can be determined in two alternative ways. 

\paragraph{From least squares variables}

Recalling the definition of the dummy variables $b_i = P_d\, d_i$, and noting that both $\{b_i\}$ and $\{d_i\}$ are available from the least squares solution of~\eqref{eq:VFleastsquares}, respectively collected in vectors $b$ and $d$, we can determine $P_d$ as the least-square solution of 
\begin{equation}
d\, P_d \approx b \quad \rightarrow \quad P_d = \frac{1}{\|d\|^2} \,d^T \cdot b.
\end{equation}

\paragraph{As periodic state bias}

From~\eqref{eq:windkesselFreqApprox} we recall that
\begin{equation}
\frac{P_d}{s}\approx P(s)-H(s)Q(s),
\end{equation}
where the approximation becomes exact at periodic state, after the transient contribution of $P_d$ has extinguished. Suppose that the periodic state holds for $t\geq t_c$. We can thus find $P_d$ as the constant value that best fits the approximation
\begin{equation}
P_d \approx p(t) - p_m(t), \quad t\geq t_c
\end{equation}
where
\begin{equation}
p_m(t)={\mathcal{L}^{-1}\{H(s)Q(s)\}}
\end{equation}
provides the output in absence of the distal pressure term. The best fit for $P_d$ is simply computed as the average
\begin{equation}
P_d=\frac{1}{K-c+1}\sum_{k=c}^K \left[ p(t_k)-p_m(t_k)\right]
\end{equation}

\subsubsection{Estimation from \emph{in vivo} measurements}\label{subsubsec:in_vivo}
When BC estimation is based on \emph{in vivo} or generally real-time measurements, the assumption of vanishing initial conditions on the data samples is not realistic. Data recording starts at some time instant $t_0$, at which $q(t_0)\neq 0$ and $p(t_0)\neq 0$. In this case, the dynamic evolution of the pressure signal for $t\geq t_0$ includes not only the zero-state response analyzed in the foregoing sections, but also some contribution from the zero-input (natural) response~\cite{bradde2021handling}. The latter is due to the nonvanishing initial conditions on the internal system states of the underlying dynamical system, which are unknown. The following derivations show how to extend the proposed algorithm to handle also this situation.

The relation between pressure $P(s)$ and flow rate $Q(s)$ at the outlet can be generalized as 
\begin{equation}\label{eq:windkesselFreqApproxInitial}
P(s) \approx  H(s)Q(s)+G(s)+\frac{P_d}{s}.
\end{equation}
where $G(s)$ represents the natural response contribution. The latter can be parameterized as
\begin{equation}
G(s)=\frac{B(s)}{s \cdot D(s)}, \quad  B(s)=r_0+\sum^n_{i=1}\dfrac{r_i}{s-a_i}
\end{equation}
based on the same starting poles $\{a_i\}$ and using the same denominator as in~\eqref{eq:modelstructure}. This choice is motivated by the well-known fact that both input-output and natural response contributions of any linear time-invariant system share the same poles.

With these definitions, condition~\eqref{eq:windkesselFreqApproxInitial} is rewritten as
\begin{multline}\label{eq:linearizationRT}
\left( \Dextnonu \right)P(s) \approx \left( \Nextnonu \right) Q(s) +\\ +P_d\left( \Dextznonu \right)+ \frac{r_0}{s}+\sum^n_{i=1}\dfrac{r_i}{s\cdot (s-a_i)},
\end{multline}
which replaces~\eqref{eq:linExpanded}. The time domain equivalent is obtained by applying the inverse Laplace transform to both sides and collecting the common terms
\begin{align}\label{eq:TDlinearizationRT}
d_0 \cdot p (t) + &\sum^n_{i=1} d_i \cdot {p}_i (t)
\approx   c_0 \cdot q(t) + \sum^n_{i=1}   c_i \cdot  {q}_i (t) + \underbrace{(P_d d_0+r_0)}_{b_0} \cdot \heaviside(t) + \sum^n_{i=1}\underbrace{(P_d d_i+r_i)}_{b_i}\cdot {\heaviside}_i(t).
\end{align}
When compared with~\eqref{eq:TDlinearization}, this expression differs only in the definition of the dummy variables $b_i$, which are nonetheless disregarded after solving the least squares problem~\eqref{eq:VFleastsquares}. For what concerns the estimation of the coefficients $\{c_i\}$ and $\{d_i\}$, the two problems~\eqref{eq:TDlinearization} and~\eqref{eq:TDlinearizationRT} are identical. Therefore, the proposed estimation algorithm can be applied without any modification and independently on the conditions of the system when the recording of the training signals begins.


\subsection{Implementation of high-order boundary conditions}\label{subsec:bc_implementation}
Once the estimation process is completed, the obtained model can used as a boundary condition in cardiovascular simulations. 
We already showed in Sec.~\ref{subsec:wk_method} how boundary conditions of order 1 can be represented as a three-element Windkessel model, and how it is possible to obtain Windkessel parameters from the general pole-residue form by means of~\eqref{eq:poleResiduesToParameters}.
For higher order BCs, different approaches can be adopted for their implementation into CFD solvers.
One approach is to transform the final model expression~\eqref{eq:TDlinearization} into an equivalent circuit by means of a synthesis process. Common techniques for equivalent circuit synthesis can be found in~\cite{grivet2015passive, antonini2003spice}.

An alternative approach consists in using directly the discretized differential equations obtained with VF as boundary conditions, without resorting to their equivalent circuit realization.
Since the poles identified by VF could be either real or complex, the general transfer function~\eqref{eq:H_high_order} can be rewritten as
\begin{equation}\label{eq:TF_real_complex}
H(s) = c_0 + \sum_{i=1}^{n_r} \frac{c_{r_{i}}}{s-a_{r_{i}}} + \sum_{i=1}^{n_c}\bigg(\frac{c_{c_{i}}}{s-a_{c_{i}}} + \frac{c_{c_{i}}^*}{s-a_{c_{i}}^*}\bigg),
\end{equation}
where the first sum includes the $n_r$ real poles, with $a_{r_{i}}, c_{r_{i}}\in \mathbb{R}$, while the second sum includes $n_c$ pairs of complex conjugate poles, with $a_{c_{i}}, c_{c_{i}}  \in \mathbb{C}$, and where the superscript $^*$ denotes the complex conjugate. Multiplying~\eqref{eq:TF_real_complex} by flow rate $q(t)$ and using the inverse Laplace transform leads to a set of differential equations, which can be cast in the following state space form for real poles
\begin{equation}\label{eq:real_poles}
\begin{cases}
\Dot{x}_i(t)&=a_{r_{i}} x_i(t) + q(t)\\
p_r(t) &= \sum_{i=1}^{n_r} c_{r_{i}} x_i(t)
\end{cases}
\end{equation}
and in the following form for complex pole pairs
\begin{equation}\label{eq:complex_poles}
\begin{cases}
\Dot{x}'_i(t)&=\sigma_{c_{i}} x'_i(t) + \omega_{c_i} x''_{i}(t) + 2q(t)\\
\Dot{x}''_{i}(t)&=-\omega_{c_i} x'_i(t) +\sigma_{c_{i}} x''_{i}(t) \\
p_c(t) &= \sum_{i=1}^{n_c}(c_{c_{i}}' x'_i(t) + c_{c_{i}}'' x''_{i}(t))
\end{cases}
\end{equation}
where $c_{c_{i}} = c_{c_{i}}' + j c_{c_{i}}''$ and $a_{c_{i}} = \sigma_{c_{i}} + j\omega_{c_{i}}$.
Systems~\eqref{eq:real_poles} and~\eqref{eq:complex_poles} can be discretized in time using the same techniques used to discretize 1D and 3D Navier-Stokes equations in CFD solvers. For example, for the implementation in the Nektar1D solver~\cite{alastruey2012arterial} where the Forward Euler method was used, the real pole states were obtained as 
\begin{equation}\label{eq:disc_x_k}
x_i(t_k) = x_i(t_{k-1}) + \Delta t\cdot [a_{r_{i}}x_i(t_{k-1}) + q(t_{k-1})].
\end{equation}
A similar relationship holds for the coupled states associated with complex pole pairs. The two sets of equations~\eqref{eq:disc_x_k} and~\eqref{eq:disc_p_k} provide the total pressure at the $k$-th time step
\begin{equation}\label{eq:disc_p_k}
p(t_k) = p_r(t_k) + p_c(t_k) + c_0 q(t_k)
\end{equation}
in terms of flow rate at present and past time steps $q(t_k)$, $q(t_{k-1})$, and instrumental state variables $x_i(t_{k-1})$, which must be stored to enable the evaluation of the recurrence relations~\eqref{eq:disc_x_k}. We remark that the above implementation provides a direct extension of the actual implementation of 3WK boundary conditions in the solver Nektar1D~\cite{alastruey2012arterial}.

Alternatively, since the proposed estimation method represents the model by means of a transfer function, the latter can be used directly into dedicated solvers for the simulation of dynamical systems, such as Simulink~\cite{documentationsimulation}, which are also widely used in cardiovascular settings. 

\section{Numerical Results} \label{sec:results}
This section provides numerical results for the experiments related to boundary conditions estimation based on Time-Domain Vector Fitting. In particular, after a general description in Sec.~\ref{subsec:setup} of the experimental setup, in Sec.~\ref{subsec:windkessel} we evaluate the ability of the proposed method to estimate the parameters of 3WK models, compared to two other methods presented in the literature, and we assess the level of accuracy obtained when these models are used as boundary conditions in place of a more detailed vascular model. Then, we quantify the sensitivity of the obtained estimates to noise in Sec.~\ref{subsec:noise} and their validity under changes of the physiological state of the patient (Sec.~\ref{subsec:mental_stress}). Lastly, in Sec.~\ref{subsec:higher} we evaluate the accuracy and robustness of the proposed algorithm for the estimation of higher order models.

\subsection{Experimental setup}\label{subsec:setup}
Experiments were conducted on a 1D arterial network representing the 55 largest arteries, as depicted in the left panel of Fig.~\ref{fig:55_artery_network}. 
One-dimensional models provide an accurate approximation of blood flow in larger arteries, as documented in~\cite{alastruey2016impact, xiao2014systematic}, with a significant reduction in the computational cost with respect to 3D fluid-structure interaction (FSI) simulations.
The parameters characterizing each segment are reported in~\cite{alastruey2012arterial}, and refer to a normotensive case. The inlet boundary condition corresponds to a realistic inlet flow at the aortic root~\cite{alastruey2012arterial}, while the outlet boundary conditions at each terminal vessel consist of a 3WK model, whose parameters are detailed in~\cite{alastruey2012arterial}. 
The 55-artery model was simulated using the Nektar1D solver~\cite{alastruey2012arterial}, which solves the nonlinear, one-dimensional blood flow equations in a given network of compliant vessels. Specifically, Nektar1D adopts the method of characteristics and the discontinuous Galerkin numerical scheme~\cite{alastruey2012arterial} to solve numerically the system of equations.
The solution provided by Nektar1D on the 55-artery network represents the reference solution for the model.

The 55-artery model was then reduced to a 21-artery model, containing only segments from the aorta up the first generation of bifurcations, by substituting the remaining segments with lumped parameter boundary conditions. A representation of the reduced model is shown on the right of Fig.~\ref{fig:55_artery_network}, where the boundary conditions are represented as 3WK models. The original network on the left was truncated at the end of segments 3 (rachiocephalic artery), 15 (left common carotid artery), 19 (left subclavian artery), 29 (celiac artery), 42 (left common iliac artery), and 43 (right common iliac artery).
The parameters of the corresponding lumped parameter terminations were estimated with the TDVF algorithm presented in Sec.~\ref{sec:methodology} through the following steps:
\begin{itemize}
	\item the 55-artery model was simulated using Nektar1D, providing the reference solution of the model;
	\item pressure and flow rate waveforms at the truncation sites were extracted from the reference solution of the 55-artery model;
	\item for each truncation location, pressure and flow rate data were fed into the TDVF algorithm, which estimated simultaneously the parameters of the lumped boundary conditions, as explained in Section~\ref{sec:methodology};
	\item the segments below the truncation site were substituted with the estimated boundary conditions;
	\item the reduced 21-artery model obtained in this way was simulated using Nektar1D.
\end{itemize}

\subsection{Estimation of Windkessel Boundary Conditions}\label{subsec:windkessel}
\begin{figure}
	\centerline{\includegraphics[width=0.49\columnwidth,clip]{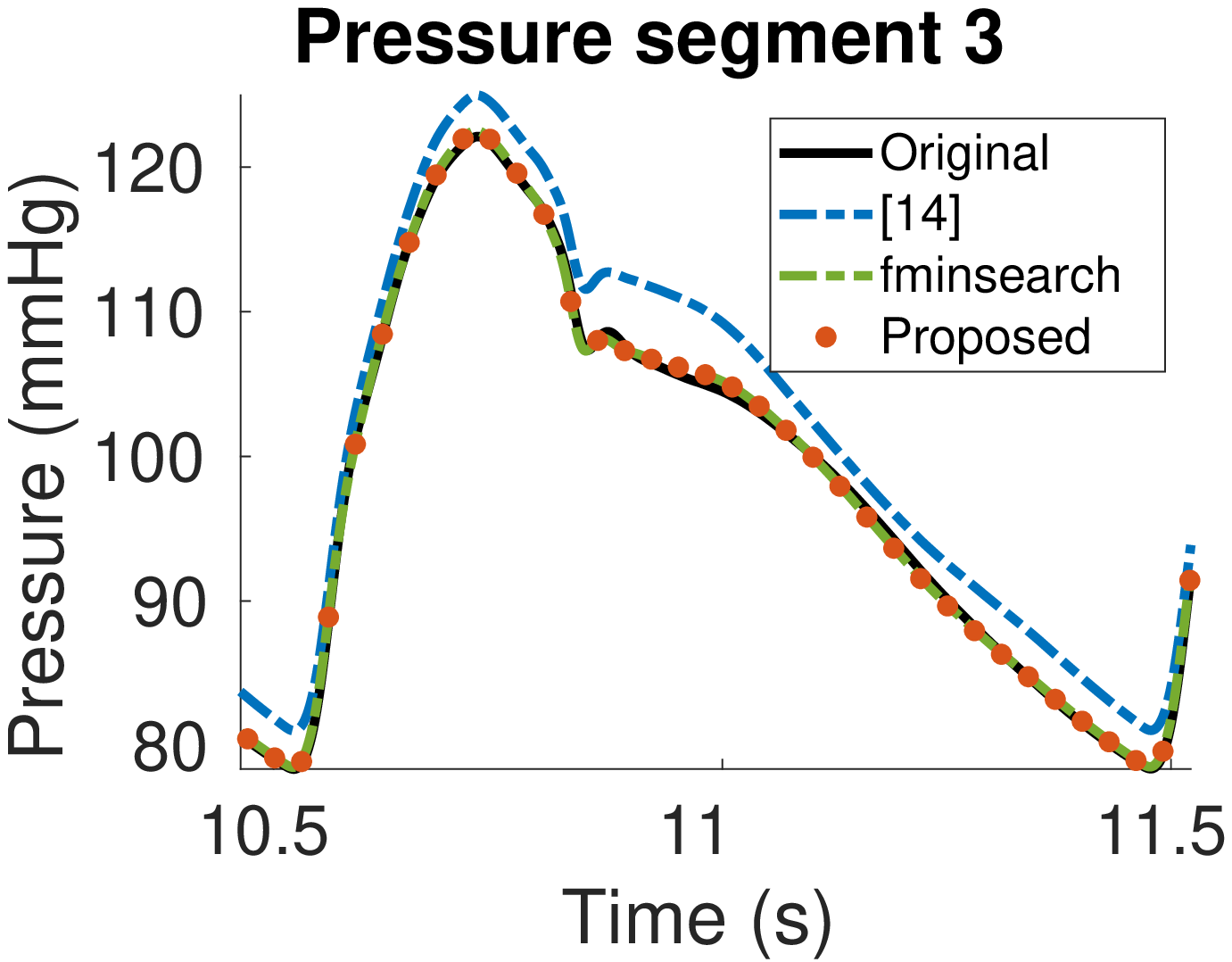}%
		\includegraphics[width=0.49\columnwidth,clip]{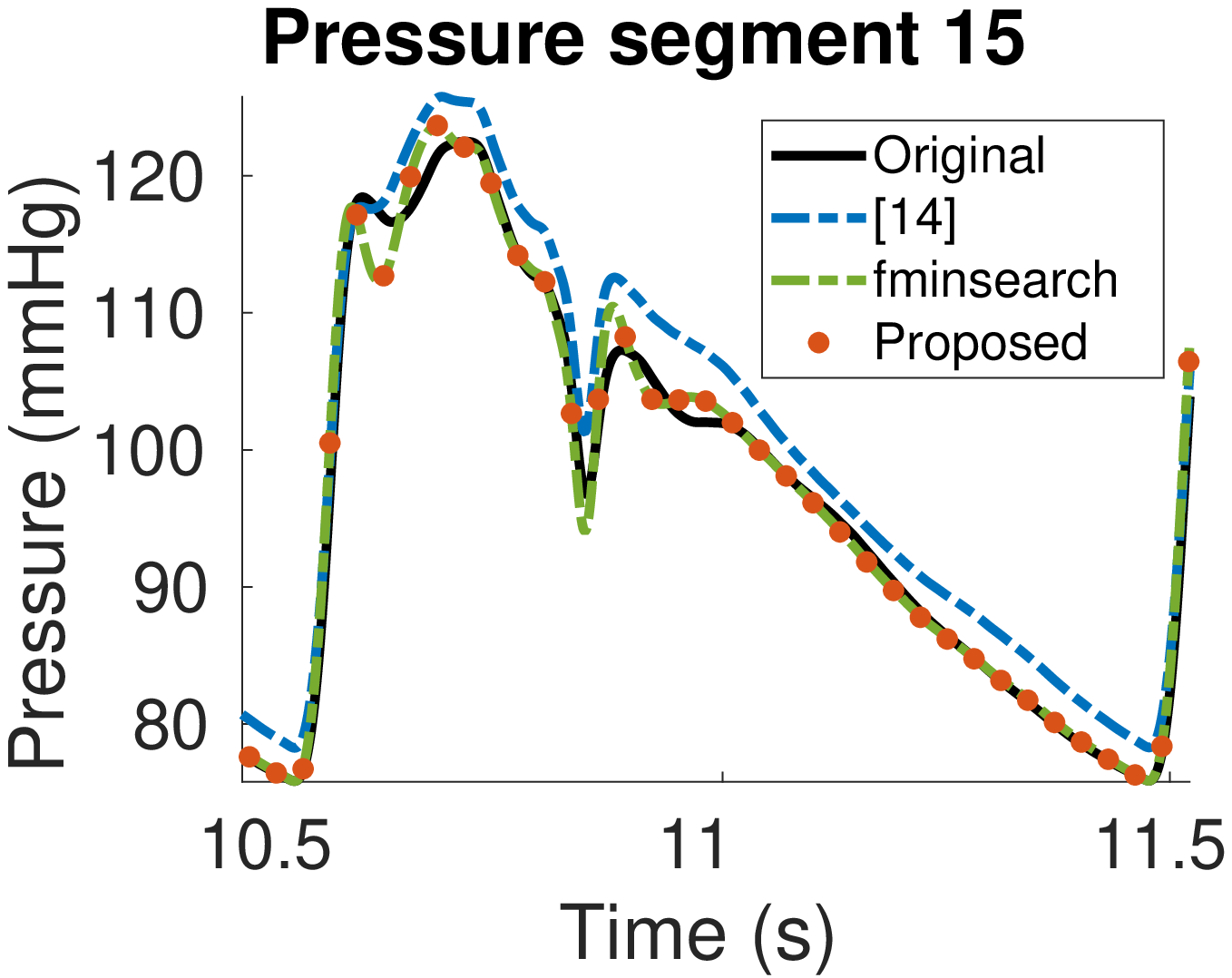}}
	\vspace*{0.2cm}
	\centerline{\includegraphics[width=0.49\columnwidth,clip]{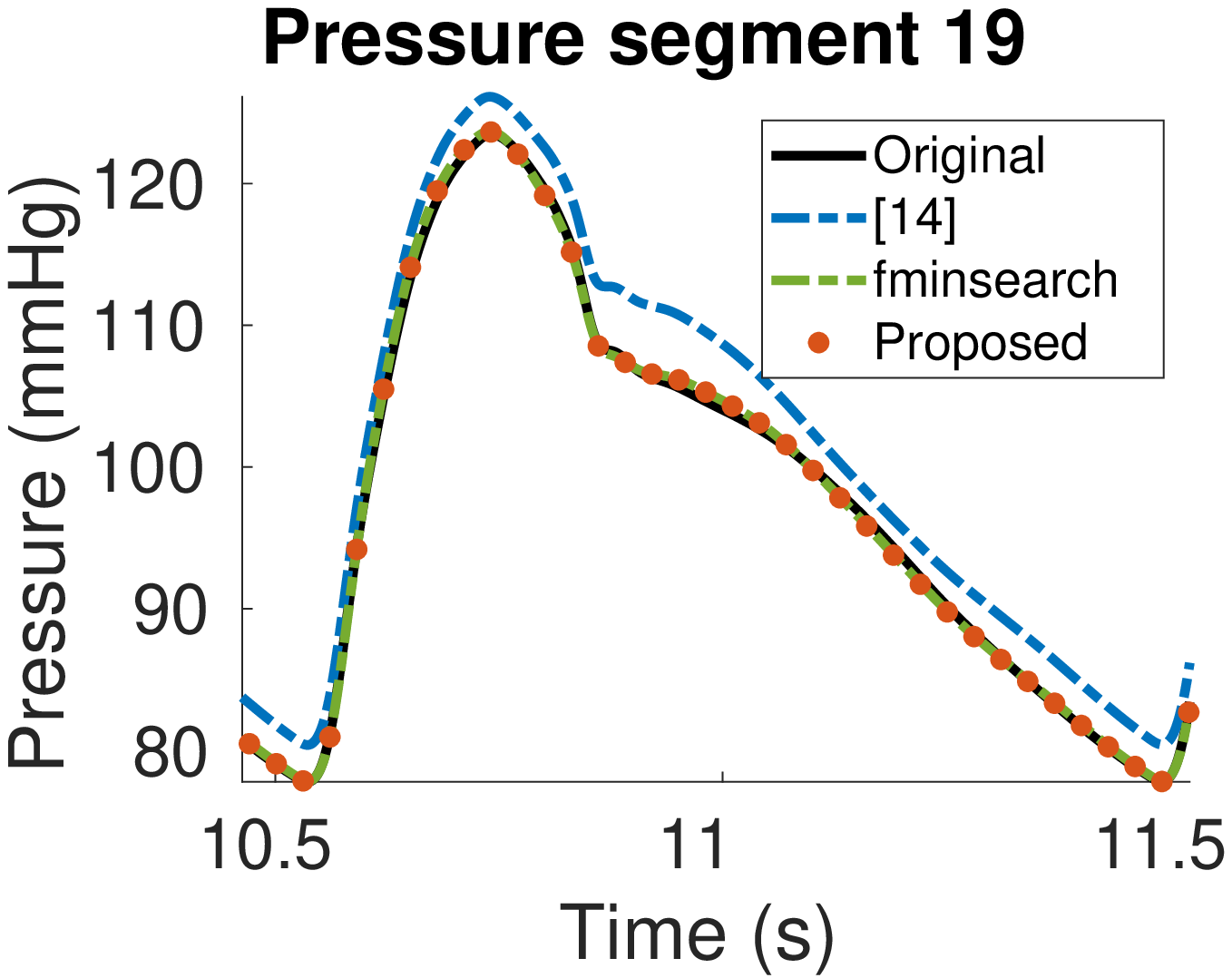}%
		\includegraphics[width=0.49\columnwidth,clip]{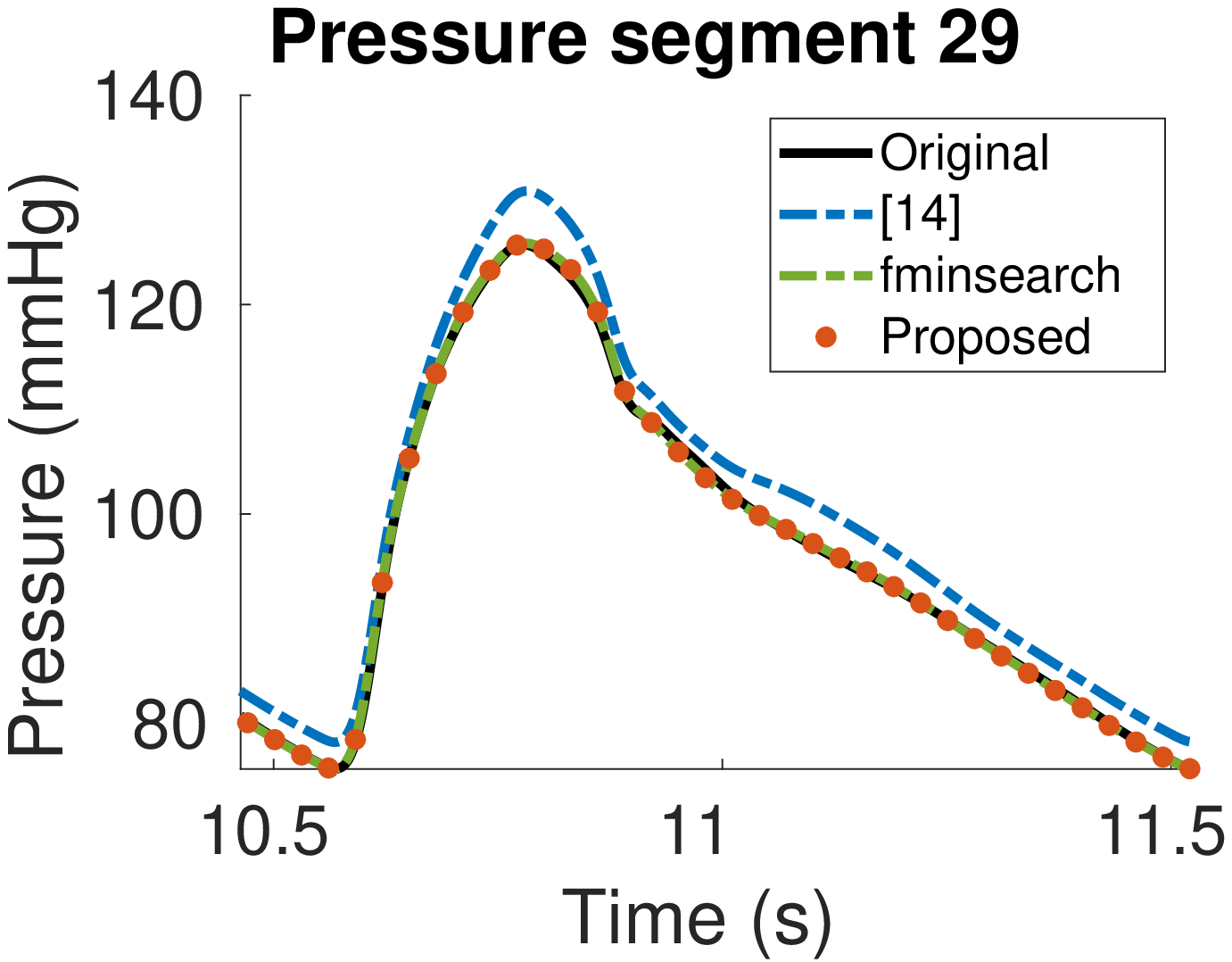}}%
	\vspace*{0.2cm}
	\centerline{\includegraphics[width=0.49\columnwidth,clip]{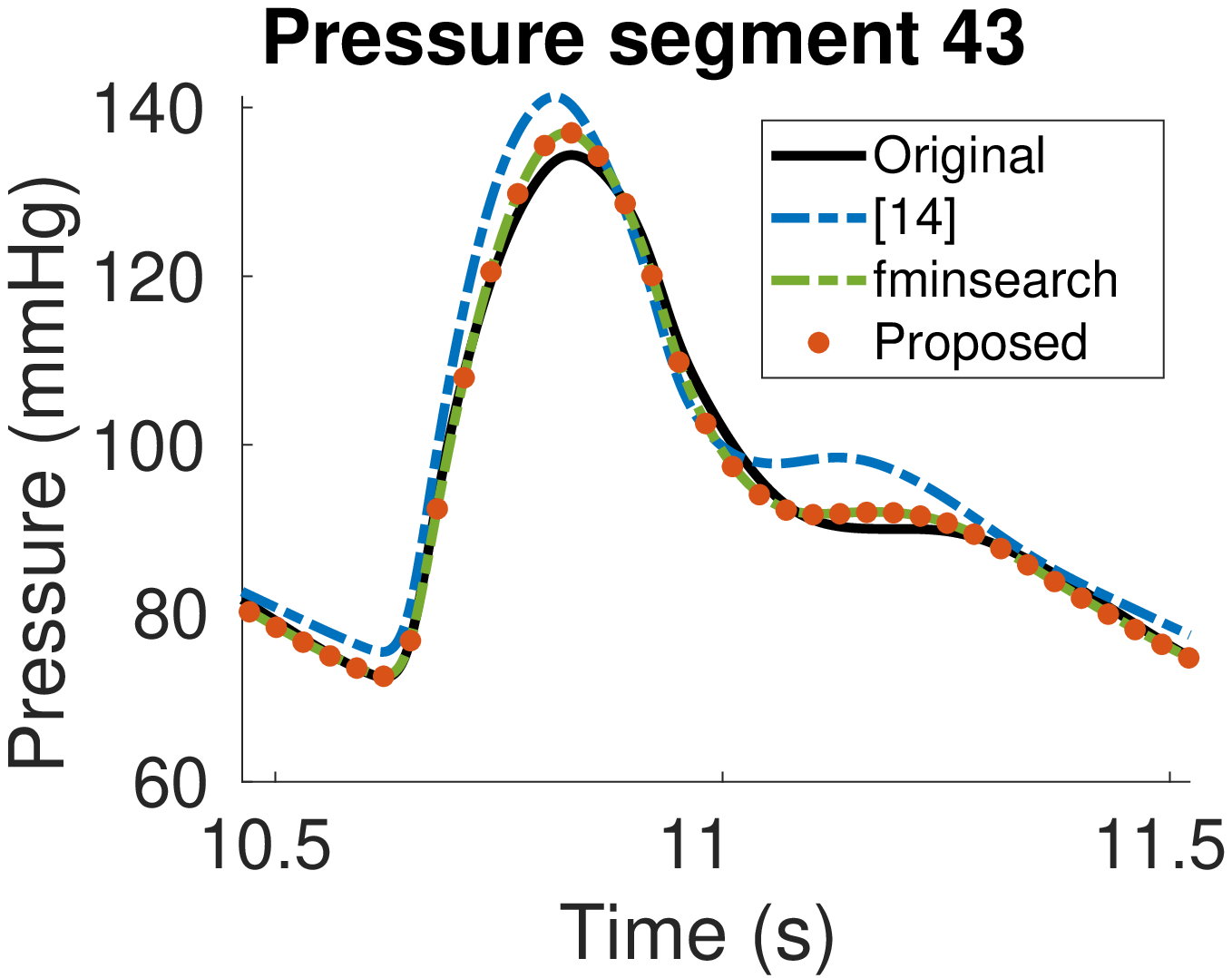}%
		\includegraphics[width=0.49\columnwidth,clip]{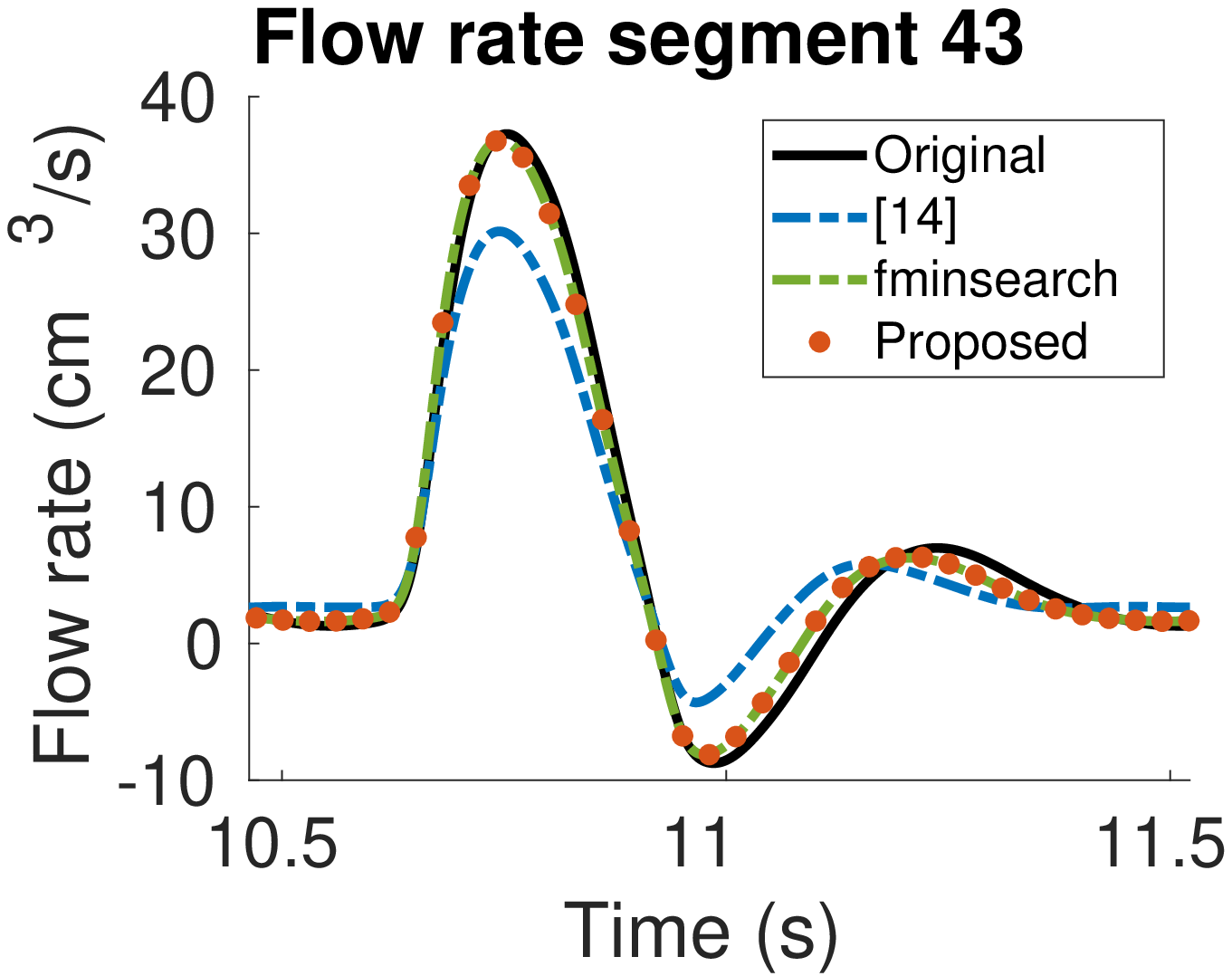}}%
	\caption{Comparison between original samples(solid black), reduction method from~\cite{epstein2015reducing} (dashed blue), \textit{fminsearch} method (dashed green), and proposed method (dashed red) for the three-element Windkessel boundary condition.} 
	\label{fig:accuracy_windkessel}
\end{figure}

The results obtained from the estimation of Windkessel parameters with TDVF have been compared to those obtained with two other methods proposed in the literature. The first was presented in~\cite{epstein2015reducing}, and selects parameters of the 3WK models such that the net resistance and total compliance of the entire system are preserved.
The second one is based on the use of the \textit{fminsearch} algorithm in MATLAB, which employs the Nelder-Mead simplex algorithm~\cite{lagarias1998convergence} to find the minimum of a given function.
In particular, the minimization problem is defined as
\begin{multline}\label{eq:fminsearch}
\min_{R_1,R_2,C,P_d} \Big\| p(t)- R_1q(t) + \frac{1}{C}\int_{0}^t e^{-\frac{1}{R_2C}(t-\tau)} q(\tau) d \tau + \frac{P_d}{R_2C}\int_{0}^t e^{-\frac{1}{R_2C}(t-\tau)}\theta(\tau) d\tau \Big\|^2
\end{multline}
Equation~\eqref{eq:fminsearch} can be derived by transforming~\eqref{eq:differentialToLaplace} back into time domain, and expressing the input $q(t)$ and the Heaviside function $\theta(t)$ by means of recursive convolutions.
The four unknown parameters $R_1$, $R_2$, $C$ and $P_d$, which were determined by means of \textit{fminsearch}, were normalized to obtain a faster convergence of the algorithm.
Fig.~\ref{fig:accuracy_windkessel} displays the obtained pressure waveforms at the truncation locations of the model, comparing the reference solution from the 55-artery model (black curve) to those from the reduced 21-artery model with 3WK parameters obtained with the technique presented in~\cite{epstein2015reducing} (dashed blue curve), with \textit{fminsearch} (dashed green curve), and with the proposed method (red dots). The curves obtained with \textit{fminsearch} and the proposed method represent the best approximation of the original responses. 
The average and maximum errors for the pressure curves displayed in Fig.~\ref{fig:accuracy_windkessel} are reported in Table~\ref{table:wk_errors}: the results obtained with \textit{fminsearch} and the proposed method are comparable in terms of accuracy, and with average errors always lower than 1.1\%, up to one order of magnitude smaller than the alternative method proposed in~\cite{epstein2015reducing}.
The latter does not provide an estimation of $P_d$, so the original value of 10 mmHg used in the 55-artery model was maintained for all outlets. This choice causes a visible offset of the obtained pressure curves with respect to the original curves, noticeable in Fig.~\ref{fig:accuracy_windkessel}, confirming the necessity to estimate $P_d$ from measurements at each truncation point, instead of setting it to a fixed value common to all outlets.
A comparison of the 3WK parameters obtained with the different methods at each truncated segment is reported in Table~\ref{table:3wk_values}.
Even if the \textit{fminsearch} method is a valid solution for estimating Windkessel parameters, its extension to higher order models is problematic, as it would require an increasing number of parameters to estimate. More importantly, the user would need to choose a representation of the model to define a suitable cost function that will be minimized, as in~\eqref{eq:fminsearch}. Using the pole-residue representation, for example, would require to know the exact number of real and complex poles beforehand. It would be even more difficult to set a specific topology for the lumped circuit, just knowing the model response.

\begin{table}
	\centering
	\caption{Approximation errors for pressure curves at truncation locations, Windkessel case (Sec.~\ref{subsec:windkessel})}\label{table:wk_errors}
	\begin{tabular}{|c|c|c|c|}
		\hline
		Segment            & Method     & Max error (\%) & Avg error (\%)\\ \hline
		\multirow{3}{*}{3} & \cite{epstein2015reducing}          &   5.1        &     3.08      \\ 
		& \textit{fminsearch}     &  0.67     &   0.30        \\
		& Proposed &        0.67   &      0.30     \\ \hline
		\multirow{2}{*}{15} &      \cite{epstein2015reducing}     &     6.51      &       3.24    \\  
		&      \textit{fminsearch}        & 4.40          &       0.89    \\ 
		& Proposed &     4.30      &    0.88       \\ \hline
		\multirow{2}{*}{19} &      \cite{epstein2015reducing}     &   4.85        &       3.15    \\ 
		&      \textit{fminsearch}        &     0.63      &      0.30     \\ 
		& Proposed &       0.62    &    0.30       \\ \hline
		\multirow{2}{*}{29} &      \cite{epstein2015reducing}     &   4.31        &    3.12       \\ 
		&      \textit{fminsearch}        &   0.77        &     0.25      \\
		& Proposed &       0.76    &   0.26        \\ \hline
		\multirow{2}{*}{42-43} &      \cite{epstein2015reducing}     &   9.16        &    4.2       \\ 
		&      \textit{fminsearch}        &    2.71       &       1.1    \\
		& Proposed &       2.81    &     1.1      \\ \hline
	\end{tabular}
\end{table}

\begin{table}
	\centering
	\caption{Comparison of estimated Windkessel parameters at the outlets of the 21-artery model (Sec.~\ref{subsec:windkessel})}\label{table:3wk_values}
	\begin{tabular}{|c|c|c|c|c|c|}
		\hline
		\multicolumn{1}{|c|}{Seg.} & \multicolumn{1}{c|}{Method} & \begin{tabular}[c]{@{}c@{}}$R_1$ \\(Pa s m$^{-3}$)\end{tabular}  &  \begin{tabular}[c]{@{}c@{}}$R_2$ \\(Pa s m$^{-3}$)\end{tabular} & \begin{tabular}[c]{@{}c@{}}$C$ \\(m$^3$ Pa$^{-1}$)\end{tabular} &  \begin{tabular}[c]{@{}c@{}}$P_d$ \\(kPa) \end{tabular}\\ \hline
		\multirow{3}{*}{3}    & \cite{epstein2015reducing} &  0.18$\cdot$10$^8$ &   9.26$\cdot$10$^8$    &  9.70$\cdot$10$^{-10}$  &  1.33 \\
		& \textit{fminsearch} & 0.27$\cdot$10$^8$& 8.46$\cdot$10$^8$ & 10.5$\cdot$10$^{-10}$ & 1.54 \\
		& Proposed  &  0.26$\cdot$10$^8$ &   8.43$\cdot$10$^8$    &  10.5$\cdot$10$^{-10}$  & 1.58  \\
		
		\hline
		\multirow{3}{*}{15}   & \cite{epstein2015reducing} &  3.60$\cdot$10$^8$ &   19.2$\cdot$10$^8$    &  1.14$\cdot$10$^{-10}$  &  1.33  \\
		&\textit{fminsearch} & 6.72$\cdot$10$^8$ & 14.2$\cdot$10$^8$ & 1.31$\cdot$10$^{-10}$ & 1.58\\
		& Proposed   &  6.55$\cdot$10$^8$ &   14.2$\cdot$10$^8$    &  1.23$\cdot$10$^{-10}$  &  1.64   \\
		
		\hline
		\multirow{3}{*}{19}   & \cite{epstein2015reducing} &  1.00$\cdot$10$^8$ &   17.0$\cdot$10$^8$    &  5.39$\cdot$10$^{-10}$  & 1.33 \\ 
		& \textit{fminsearch} & 0.67$\cdot$10$^8$ & 15.5$\cdot$10$^8$ & 6.17$\cdot$10$^{-10}$ & 1.59 \\
		& Proposed   &  0.67$\cdot$10$^8$ &   15.4$\cdot$10$^8$    &  6.13$\cdot$10$^{-10}$  & 1.68\\
		
		\hline
		\multirow{3}{*}{29} & \cite{epstein2015reducing}  &  1.62$\cdot$10$^8$ &   7.58$\cdot$10$^8$    &  3.06$\cdot$10$^{-10}$  & 1.33\\
		& \textit{fminsearch} & 1.99$\cdot$10$^8$ & 6.90$\cdot$10$^8$ & 4.36$\cdot$10$^{-10}$ & 1.41 \\
		& Proposed    &  1.99$\cdot$10$^8$ &   6.91$\cdot$10$^8$    &  4.36$\cdot$10$^{-10}$  &  1.41 \\
		
		\hline
		\multirow{3}{*}{42-43}  &\cite{epstein2015reducing}  &  1.57$\cdot$10$^8$ &   14.8$\cdot$10$^8$    &  5.04$\cdot$10$^{-10}$  & 1.33     \\
		& \textit{fminsearch} & 0.98$\cdot$10$^8$ & 13.6$\cdot$10$^8$ & 6.11$\cdot$10$^{-10}$ & 1.61 \\
		& Proposed   &  0.97$\cdot$10$^8$ &  13.5$\cdot$10$^8$    &  6.06$\cdot$10$^{-10}$  & 1.71    \\
		
		\hline
	\end{tabular}
\end{table}

\subsection{Sensitivity to noise}\label{subsec:noise}
\begin{figure}
	\centerline{\includegraphics[width=0.7\columnwidth,clip]{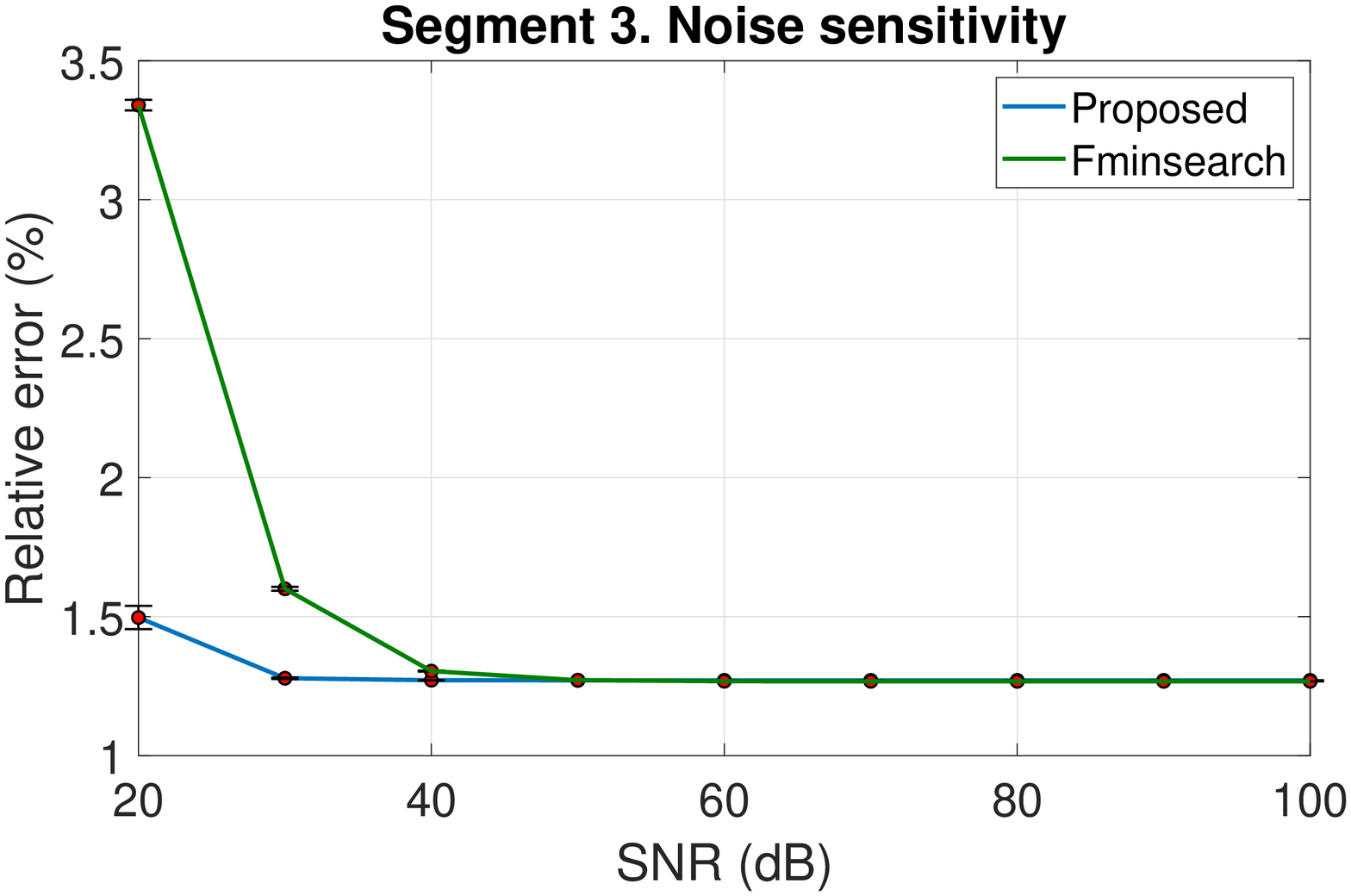}}%
	\centerline{\includegraphics[width=0.7\columnwidth,clip]{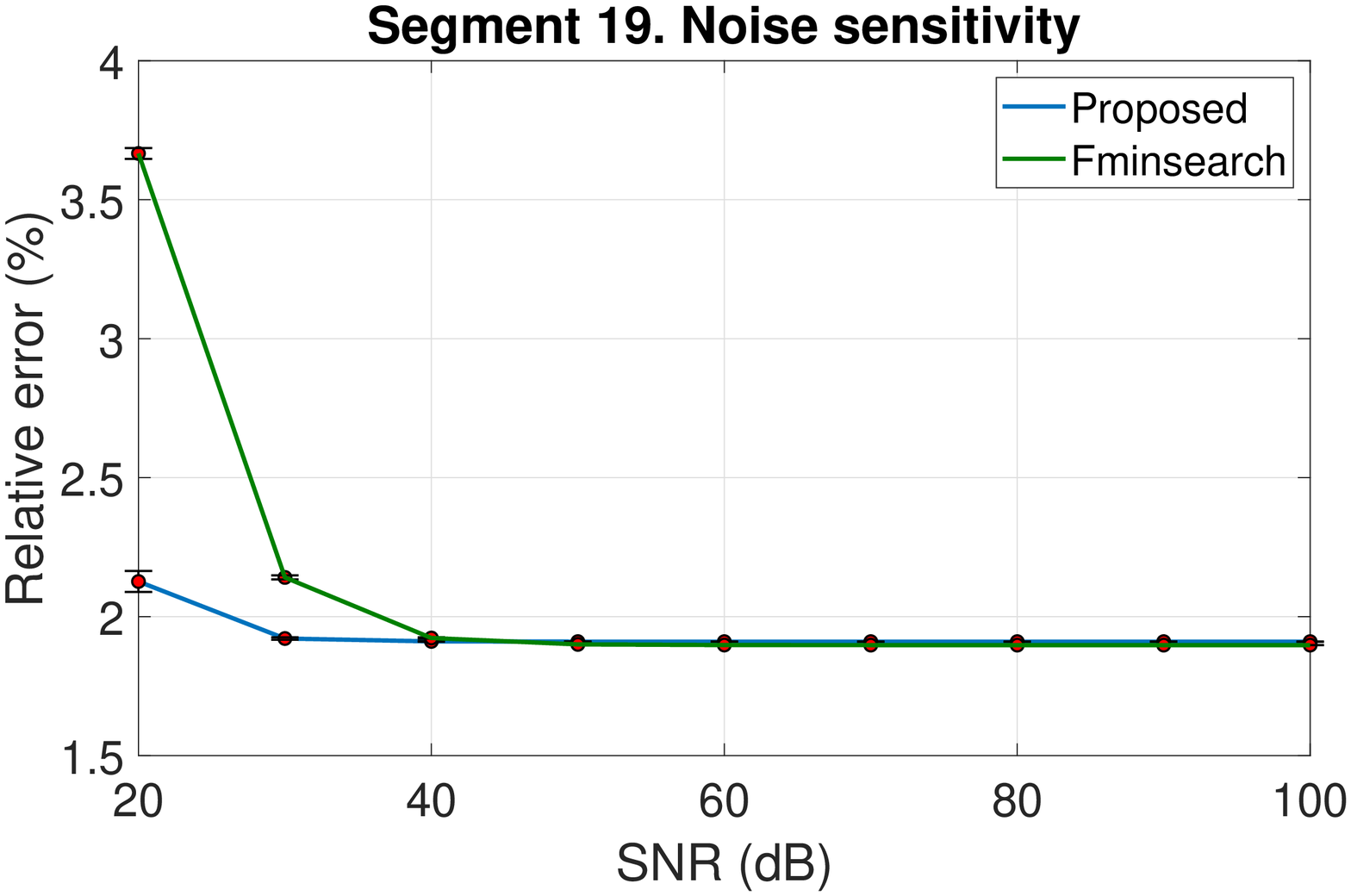}}%
	\caption{Relative error between pressure samples obtained from the original 55-artery network and Windkessel models estimated from noisy data, with different SNR values. The Windkessel parameters were estimated with TDVF (blue curve) and \textit{fminsearch} (green curve). Vertical bars in correspondence of the different SNR values indicate the standard deviation of the absolute error.}
	\label{fig:noise}
\end{figure}

To investigate the robustness of the proposed algorithm, both pressure and flow rate data were corrupted with zero-mean white Gaussian noise with signal-to-noise ratio (SNR) ranging from  20 dB up to 100 dB, corresponding to a noise standard deviation ranging between 3.95 mmHg and 3.90$\cdot 10^{-4}$ mmHg for pressure and 1.18 cm$^3$/s and 1.15$\cdot 10^{-4}$ cm$^3$/s for flow rate, respectively. For each SNR level, we generated 50 different noise realizations to corrupt the data. Then, for each corrupted dataset a 3WK boundary condition was estimated, both with the proposed method and \textit{fminsearch}. 
The results for this analysis are reported in Fig.~\ref{fig:noise}, where the average of the absolute error between the pressure samples from the 55-artery network and the output of the Windkessel models estimated at different SNR values. Moreover, at each SNR value, the bar indicates the standard deviation.
Both techniques are able to estimate the correct boundary conditions starting from data samples with SNR ranging from 100 dB down to 40 dB, without any loss of accuracy. For segment $19$, the error at 20 dB and 30 dB levels is about $2.1\%$ and $1.9\%$ respectively, while for segment $3$ these errors are $1.5\%$ and $1.3\%$, with the proposed approach performing slightly better that \textit{fminsearch}.
These results verify the robustness of the proposed estimation also in presence of noisy data, a condition more likely to occur when using patient-specific measurements instead of simulation results to drive the boundary conditions estimation.

\subsection{Validity of BCs estimated with Vector Fitting in case of mental stress}\label{subsec:mental_stress}
\begin{figure}
	\centerline{\includegraphics[width=0.49\columnwidth,clip]{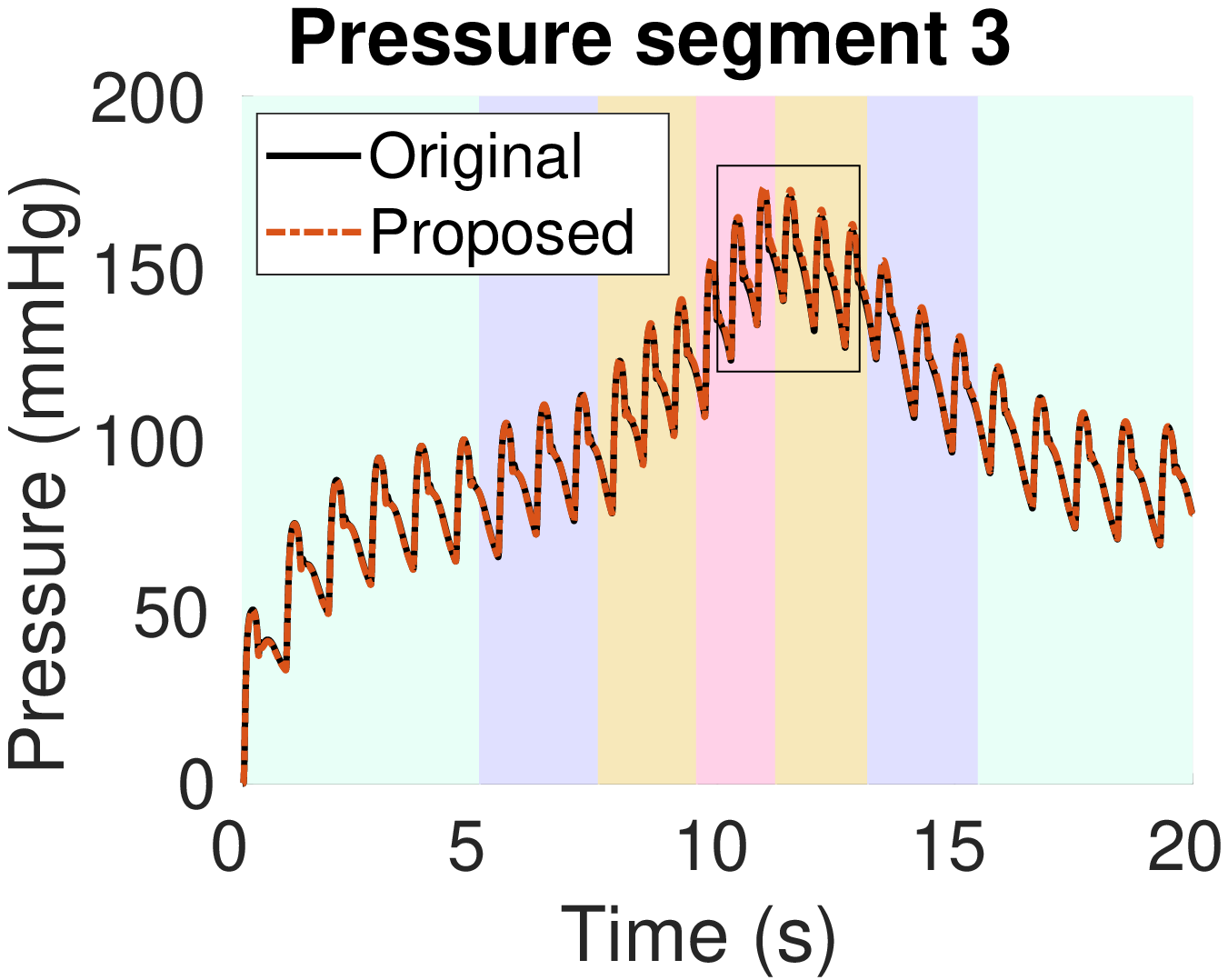}%
		\includegraphics[width=0.49\columnwidth,clip]{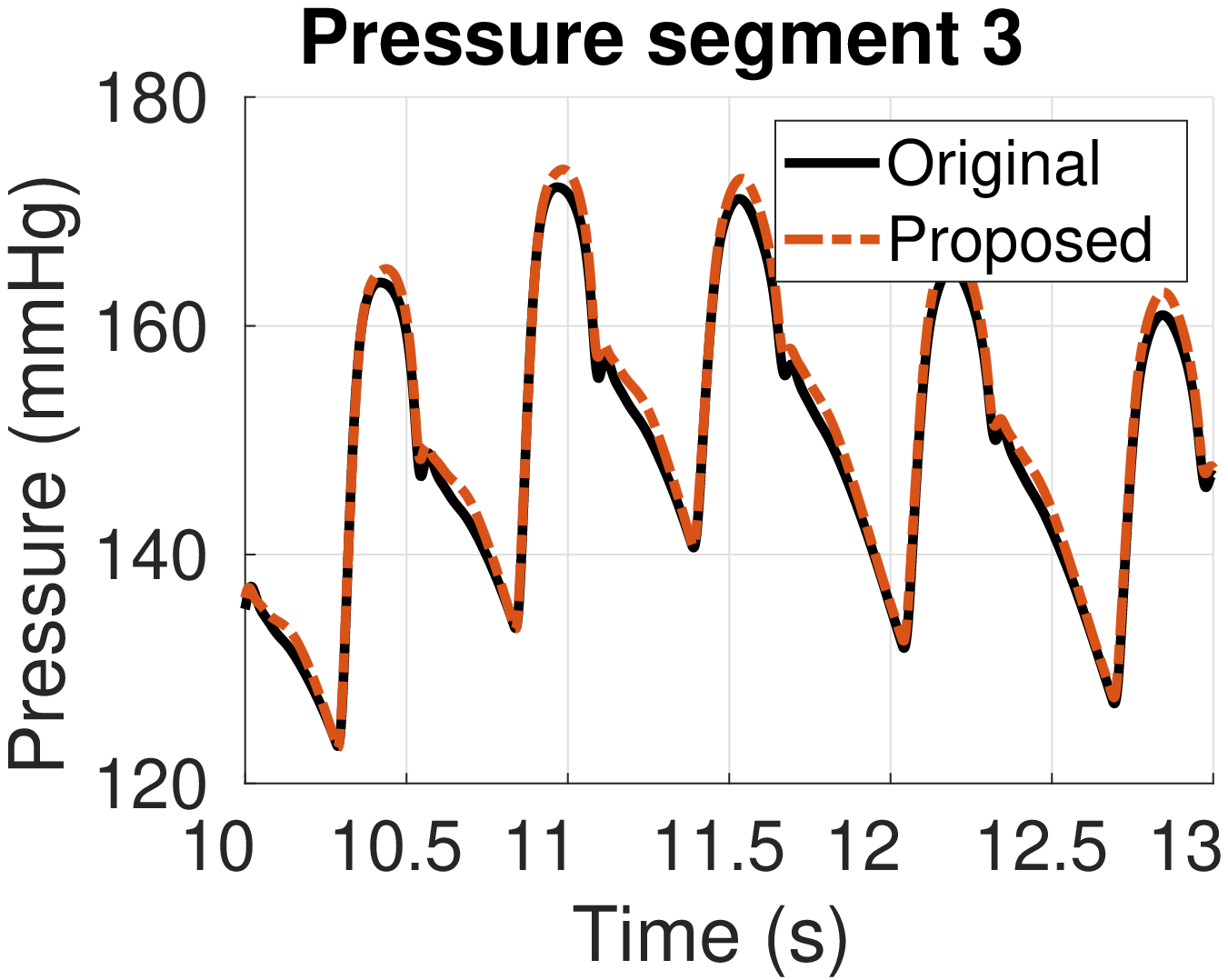}}
	\vspace*{0.2cm}
	\centerline{\includegraphics[width=0.49\columnwidth,clip]{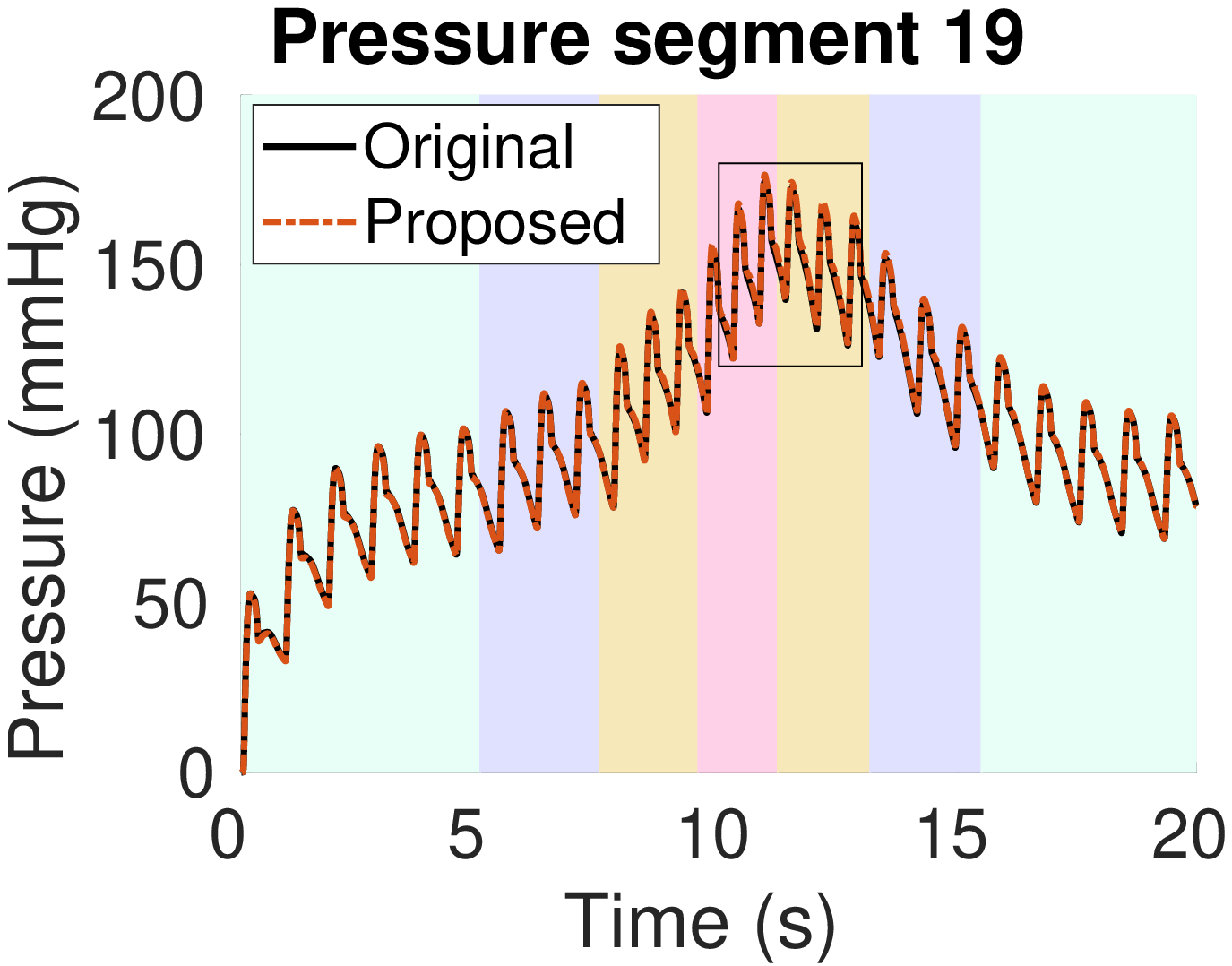}%
		\includegraphics[width=0.49\columnwidth,clip]{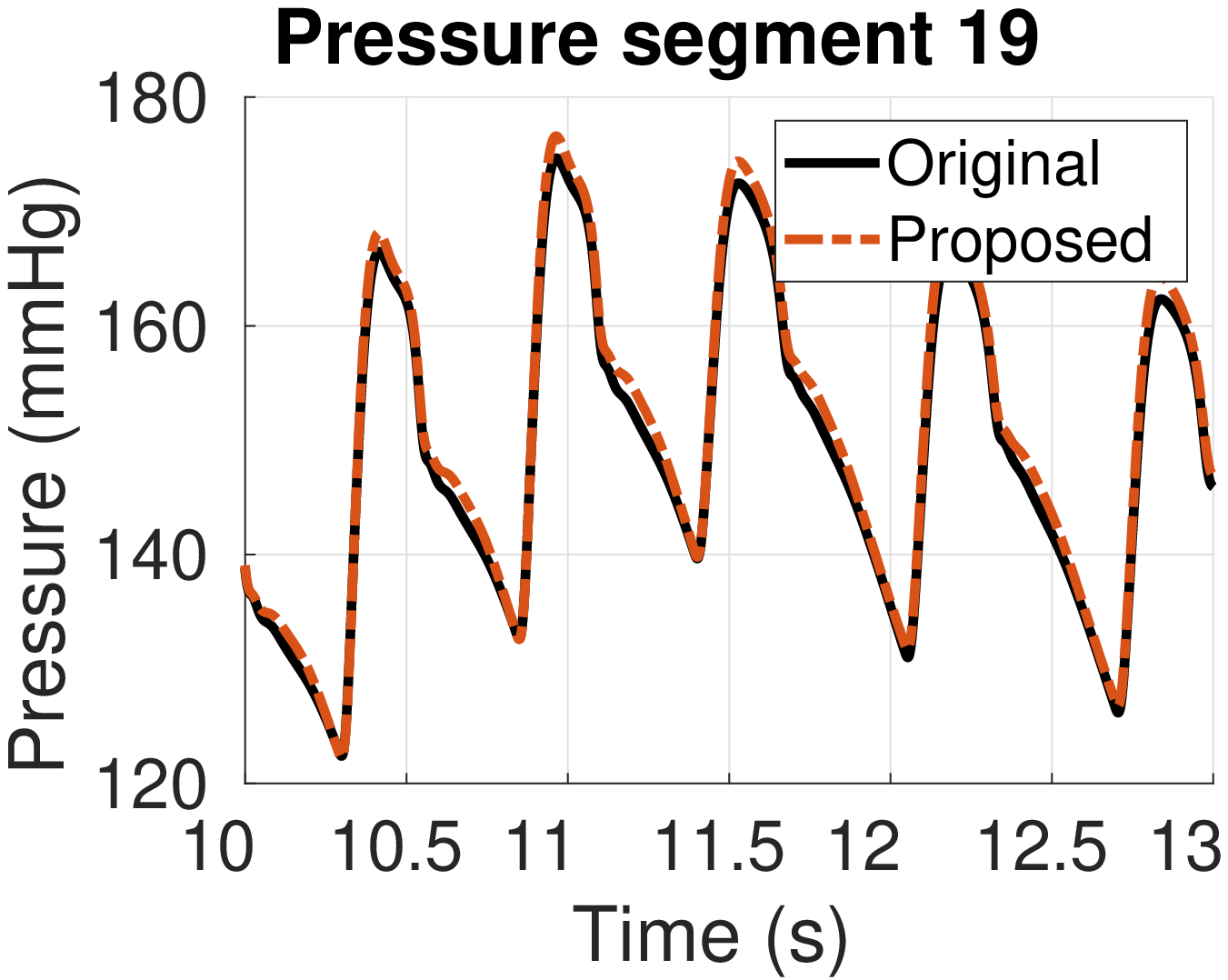}}%
	\caption{Pressure waveforms at segment 3 (top) and 19 (bottom) for the case of mental stress. Background color indicates the level of stress: relaxation (light blue), baseline (purple), medium stress (orange), high stress (pink). The plots on the right zoom on the black rectangle displayed on the plots on the left.}
	\label{fig:mental_stress}
\end{figure}

In the previous section, boundary conditions were estimated from data coming from a simulation of the cardiovascular system under normal conditions. However, under certain circumstances like physical exercise, the cardiovascular system does not operate under normal conditions anymore, experiencing physiological changes in heart rate and cardiac output. These conditions can be modeled by properly changing the flow rate at the aortic root, which is the input imposed on the 55-artery model used in this work. Therefore, it is important to verify that the boundary conditions estimated with the standard input flow rate are still valid in presence of physiological changes of the system. In order to do so, we emulated a realistic variation of the input aortic flow rate under mental stress conditions by using the dataset presented in~\cite{charlton2018assessing}, which provides different aortic root flow rates corresponding to different levels of mental stress in a human subject. These conditions translate into increased peak velocity and acceleration due to the increase in ejection fraction during stress~\cite{celka2020influence}.
An input flow with varying levels of mental stress was then generated and used as input for both the 55-artery model, chosen as a reference, and the 21-artery one. In the latter, the 3WK BCs previously estimated with the proposed approach in the normotensive case, and reported in Table~\ref{table:3wk_values}, were used. Pressure waveforms at different points of the model are reported in Fig.~\ref{fig:mental_stress}, where the results in the reference 55-artery model (black line) are compared to those in the 21-artery model. The background colors in the left panel of Fig.~\ref{fig:mental_stress} indicate the corresponding level of mental stress induced by the input aortic flow rate, varying from a relaxed state (light blue), to the baseline (purple), medium (orange) and high (pink) levels of mental stress. The corresponding heart rate and cardiac output associated to each stress level can be found in~\cite{charlton2018assessing}. From Fig.~\ref{fig:mental_stress} it is possible to see that the reduced model is able to closely follow the changes caused by the varying input flow, with average relative errors smaller than 0.7\% for both segments. The results confirm that the estimated boundary conditions are valid also in the case of a physiological change of the input flow rate. 

\subsection{Higher-order boundary conditions}\label{subsec:higher}
\begin{figure}
	\centerline{\includegraphics[width=0.49\columnwidth,clip]{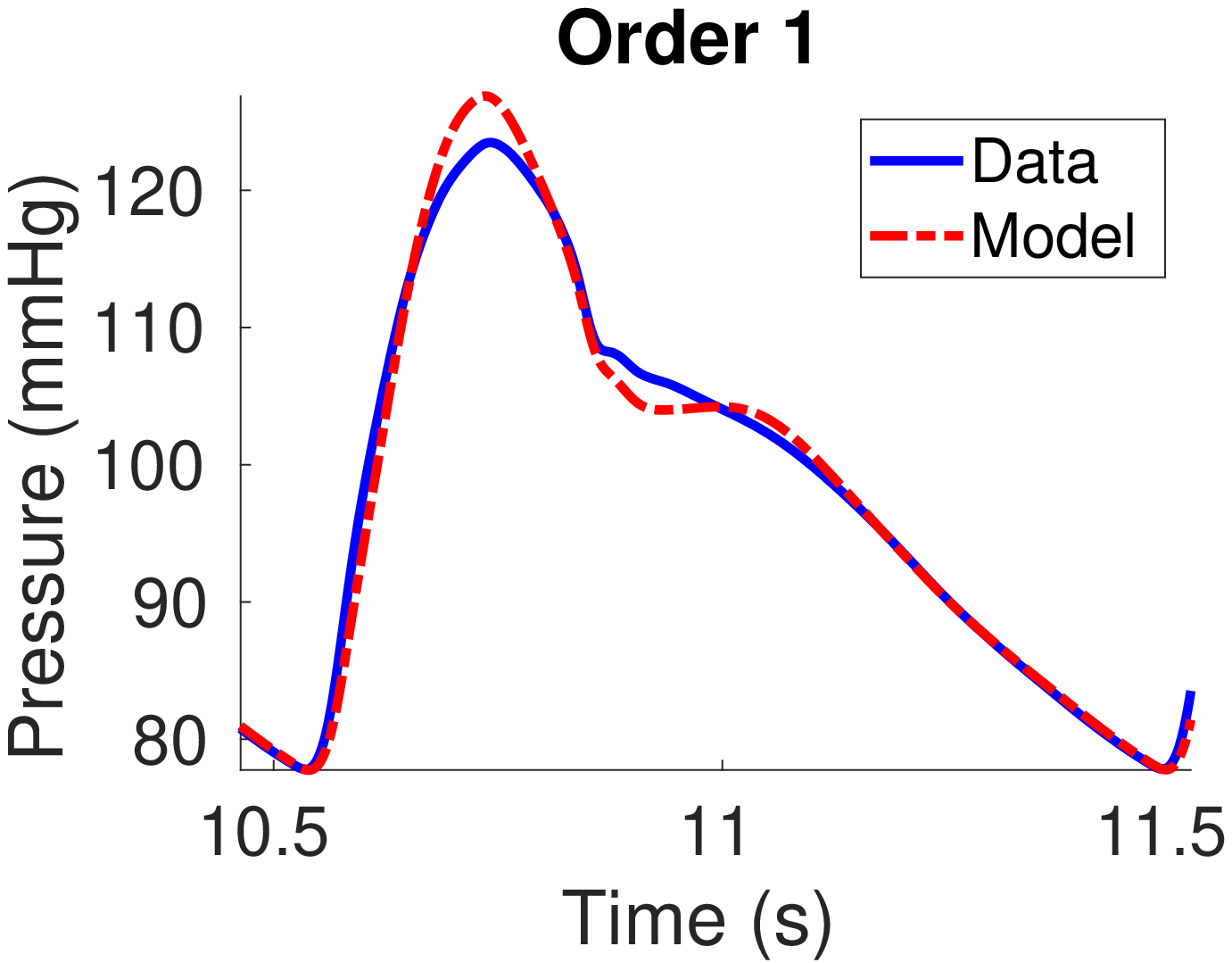}%
		\includegraphics[width=0.49\columnwidth,clip]{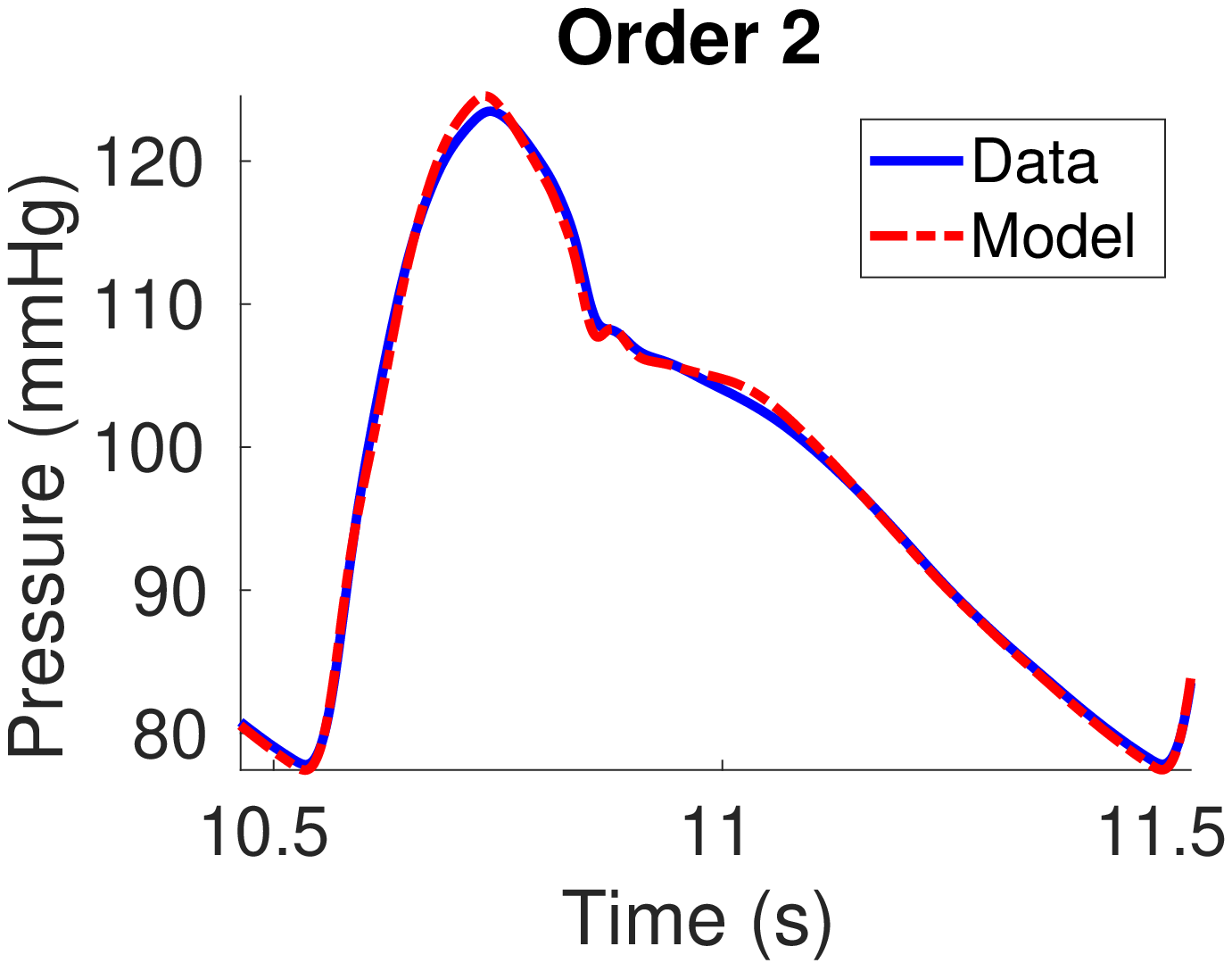}}
	\vspace*{0.2cm}
	\centerline{\includegraphics[width=0.49\columnwidth,clip]{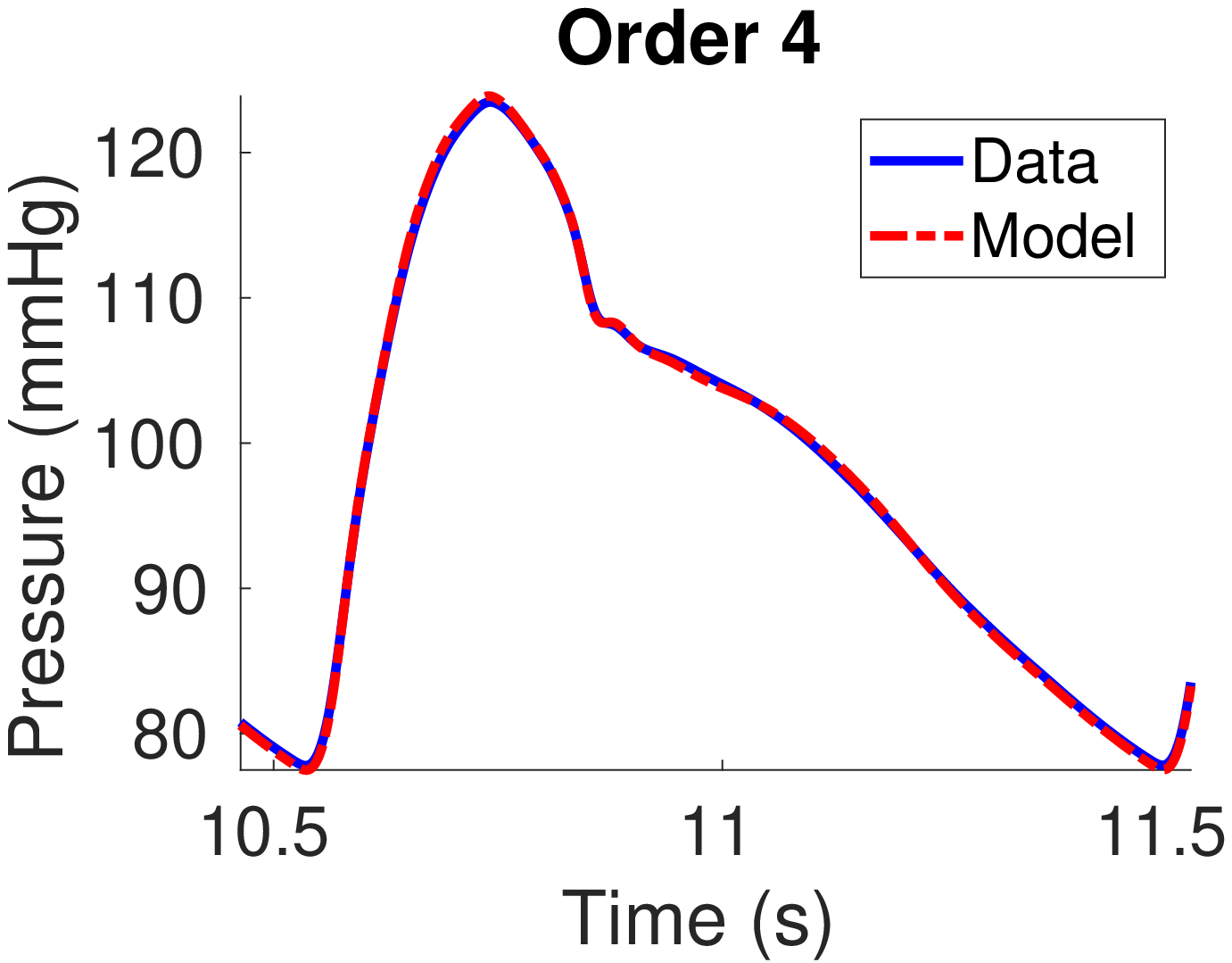}%
		\includegraphics[width=0.49\columnwidth,clip]{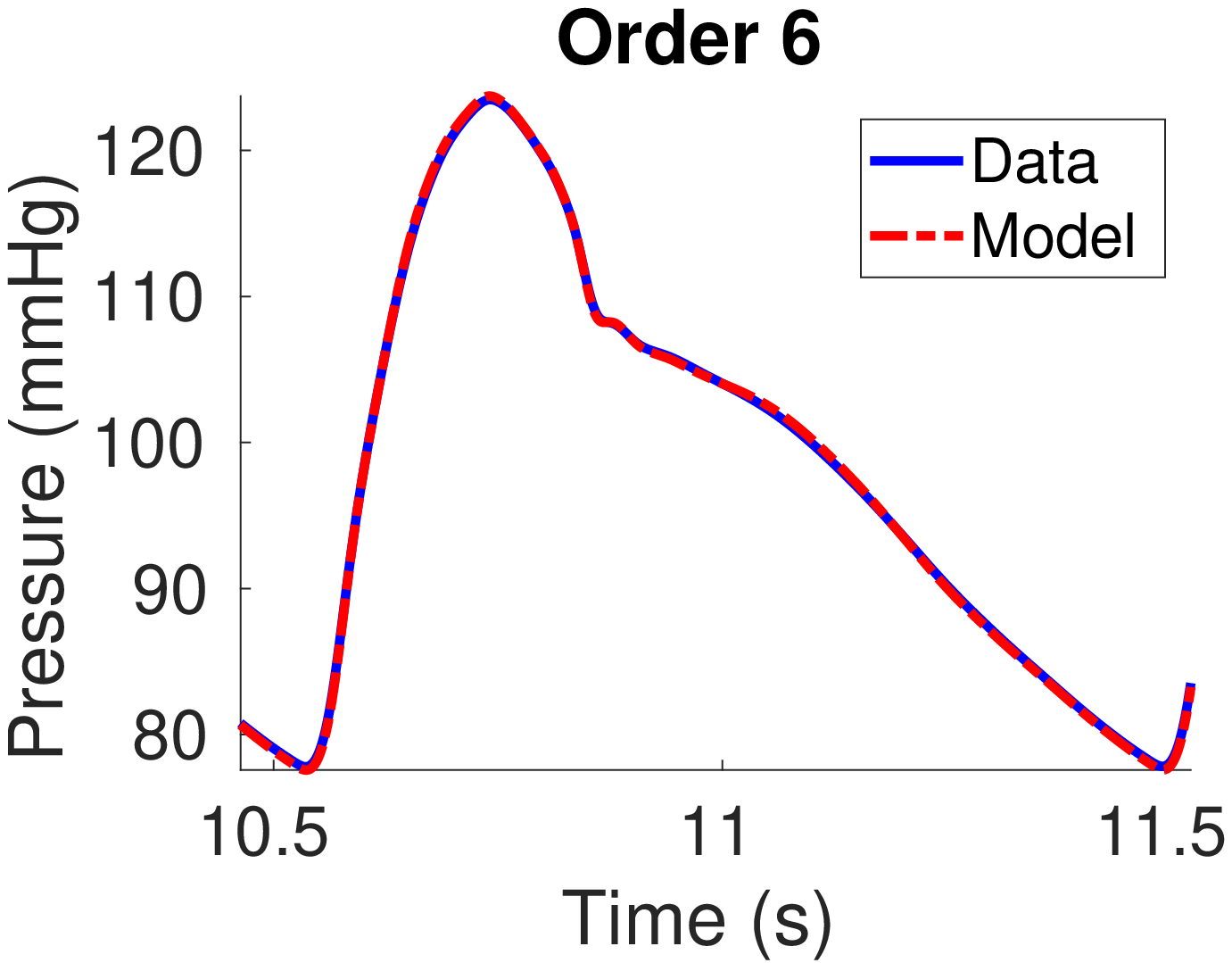}}%
	\vspace*{0.2cm}
	\centerline{\includegraphics[width=0.49\columnwidth,clip]{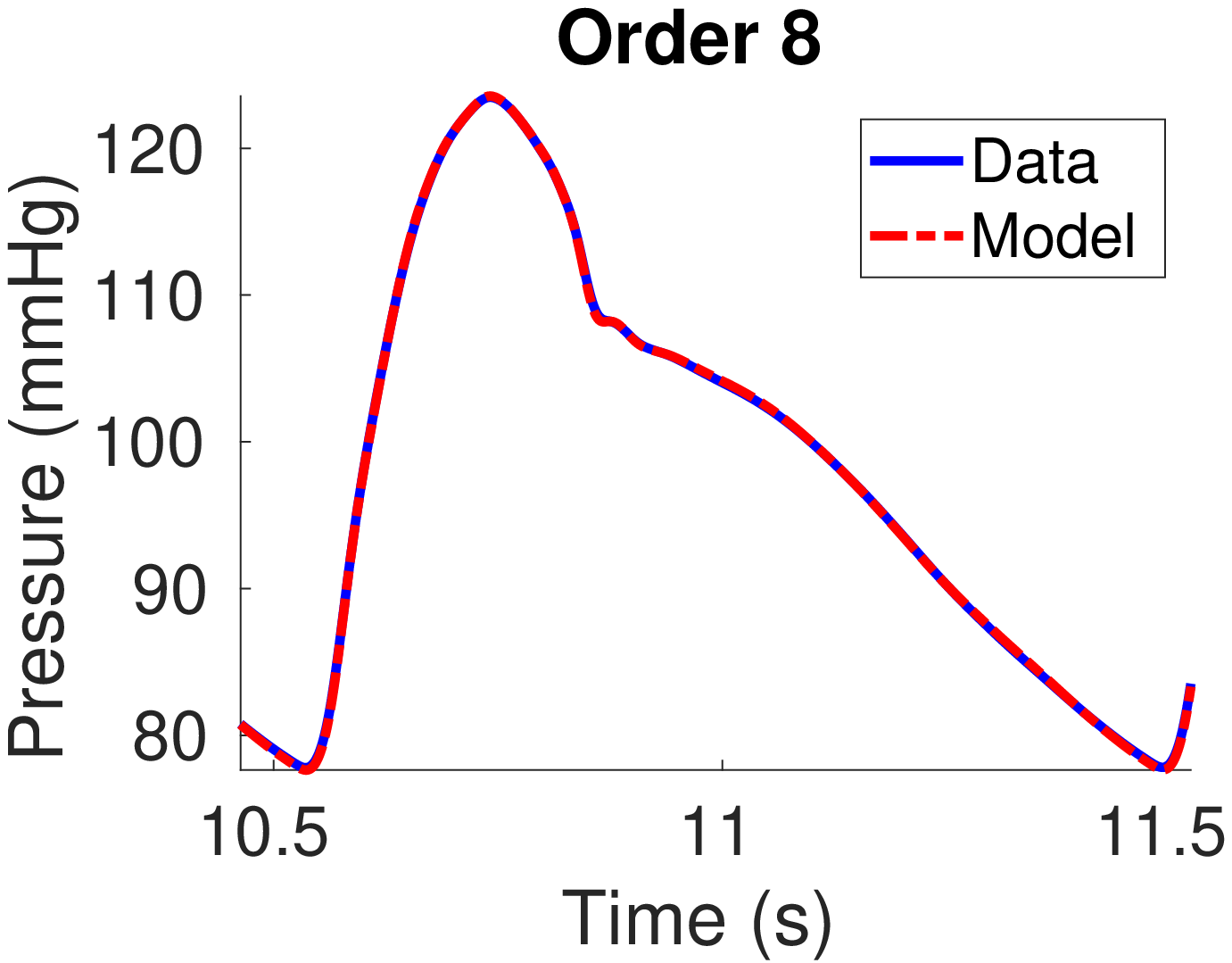}%
		\includegraphics[width=0.49\columnwidth,clip]{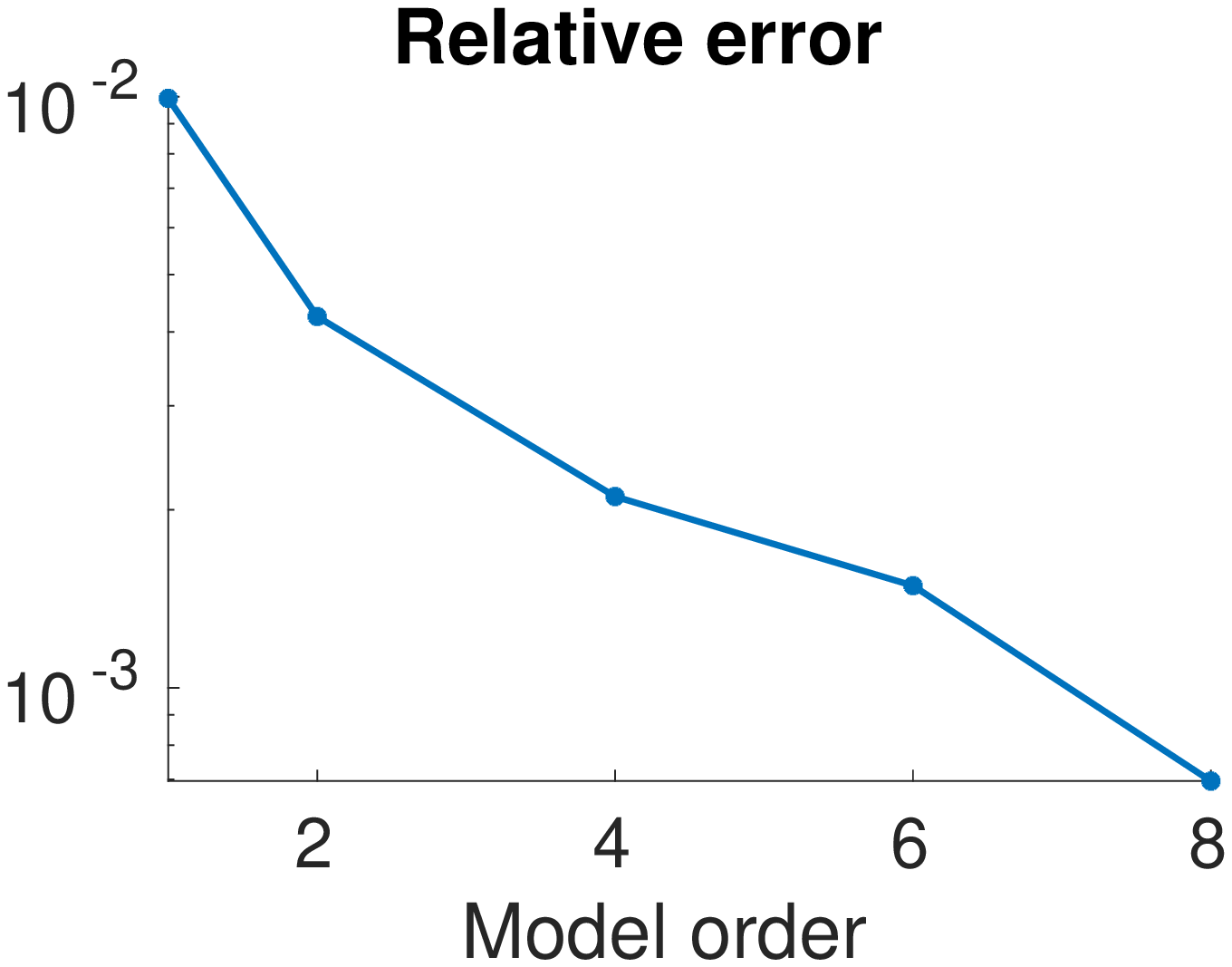}}%
	\caption{Comparison of pressure waveform in segment 19 (blue curve) against models with different order obtained with Vector Fitting. In the last plot (third row, right panel) relative error on pressure vs model order.}\label{fig:vf_higher_order}
\end{figure}

In this section, we test the use of the proposed technique for the estimation of higher order boundary conditions and we investigate their accuracy compared to standard 3WK models.
The same experimental setup presented in~\ref{subsec:setup}, consisting of the reference 55-artery model and the reduced 21-artery model, was adopted. 
In Fig.~\ref{fig:vf_higher_order}, we first compare the reference pressure from the 55-artery model used for the estimation (blue line), to the pressure estimated by the proposed model (dashed red line), for the same flow rate coming from the 55-artery model. The results reported in Fig.~\ref{fig:vf_higher_order} refer to the pressure in segment 19 fitted with models of order up to 8 (comparable results were obtained for the other segments).
It is clear from Fig.~\ref{fig:vf_higher_order} that accuracy greatly improves by increasing the model order. The right panel on the third line of Fig.~\ref{fig:vf_higher_order} shows the average relative error on pressure versus model order, suggesting a decrease of around one order of magnitude going from order 1 to order 8.
We then used the estimated models as boundary conditions for the reduced 21-artery model, simulated with Nektar1D. This step required a modification of the solver to accept boundary conditions defined as in~\ref{sec:w_high_order}, that we performed as discussed in Sec.~\ref{subsec:bc_implementation}. Pressure and flow rate curves up to order 4 at the truncated segments are displayed in Fig.~\ref{fig:accuracy_higher_order}, while the relative errors on pressure and flow rate waveforms up to order 8 are reported in Table~\ref{table:avg_errors_p_higher} and Table~\ref{table:avg_errors_q_higher} (average error), and in Table~\ref{table:max_errors_p_higher} and Table~\ref{table:max_errors_q_higher} (maximum error).
It is clear, both from plots and from numerical results, that higher order boundary conditions can model pressure and flow rate more accurately than a simple Windkessel (corresponding to order 1). In particular, a significant improvement can be seen with order 2 and order 4, where average errors can decrease up to one order of magnitude with respect to BCs of order 1. Orders above 4, instead, did not seem to provide an improvement in terms of accuracy.
Segment 15 is the only case which does not seem to benefit from higher order boundary conditions, with the error remaining nearly constant for both pressure and flow rate, even for higher orders. Looking at the corresponding plots in Fig.~\ref{fig:accuracy_higher_order}, it can be noticed that the curves obtained after the truncation are qualitatively different from the original pressure and flow in 55-artery model (black curve). A possible cause could be the higher wall viscosity of segment 15 with respect to the other terminal segments, which could increase the presence of nonlinear effects, hard to model with a linear boundary condition. However, no conclusive explanation was reached.

\begin{figure}
	\centerline{\includegraphics[width=0.49\columnwidth,clip]{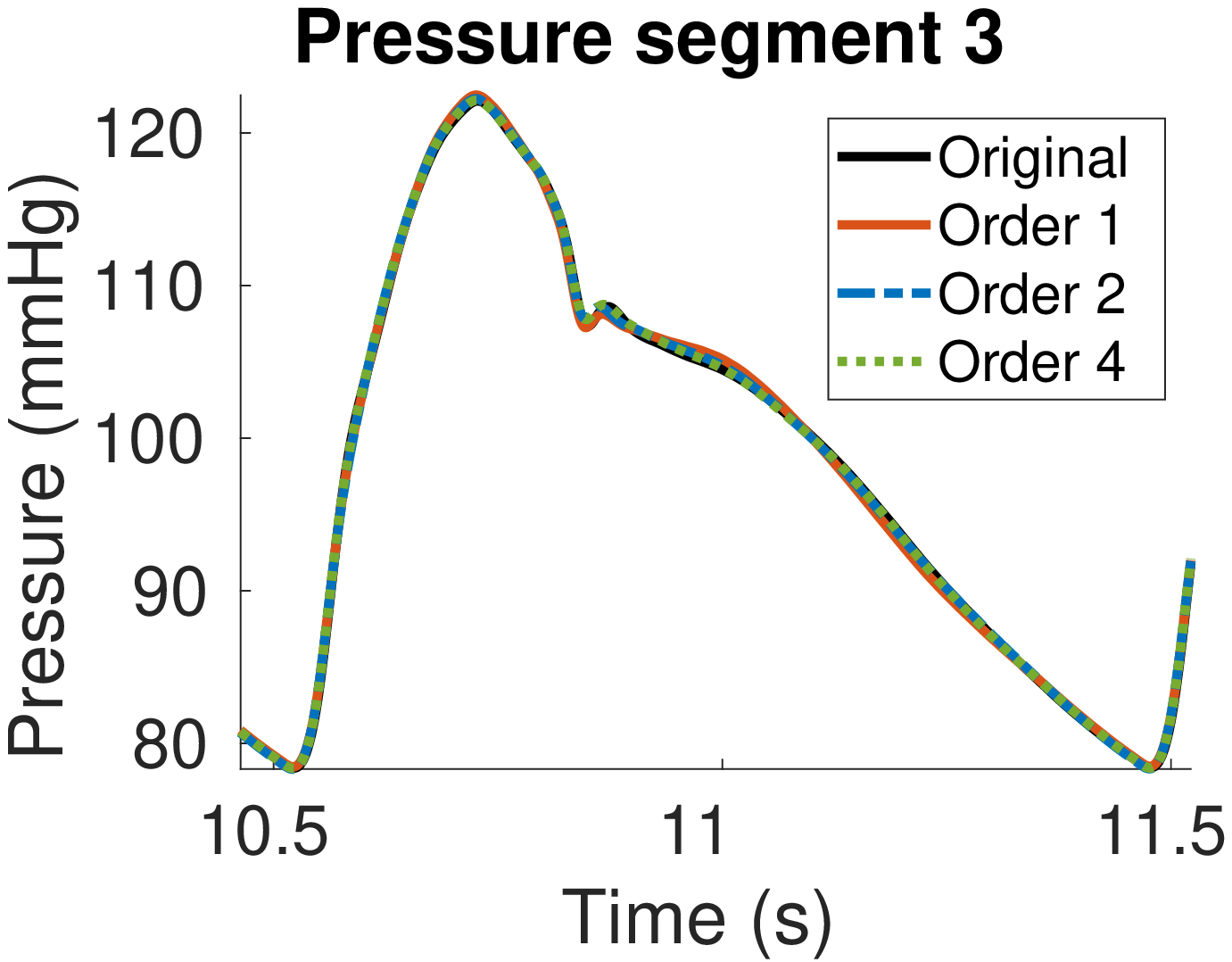}%
		\includegraphics[width=0.49\columnwidth,clip]{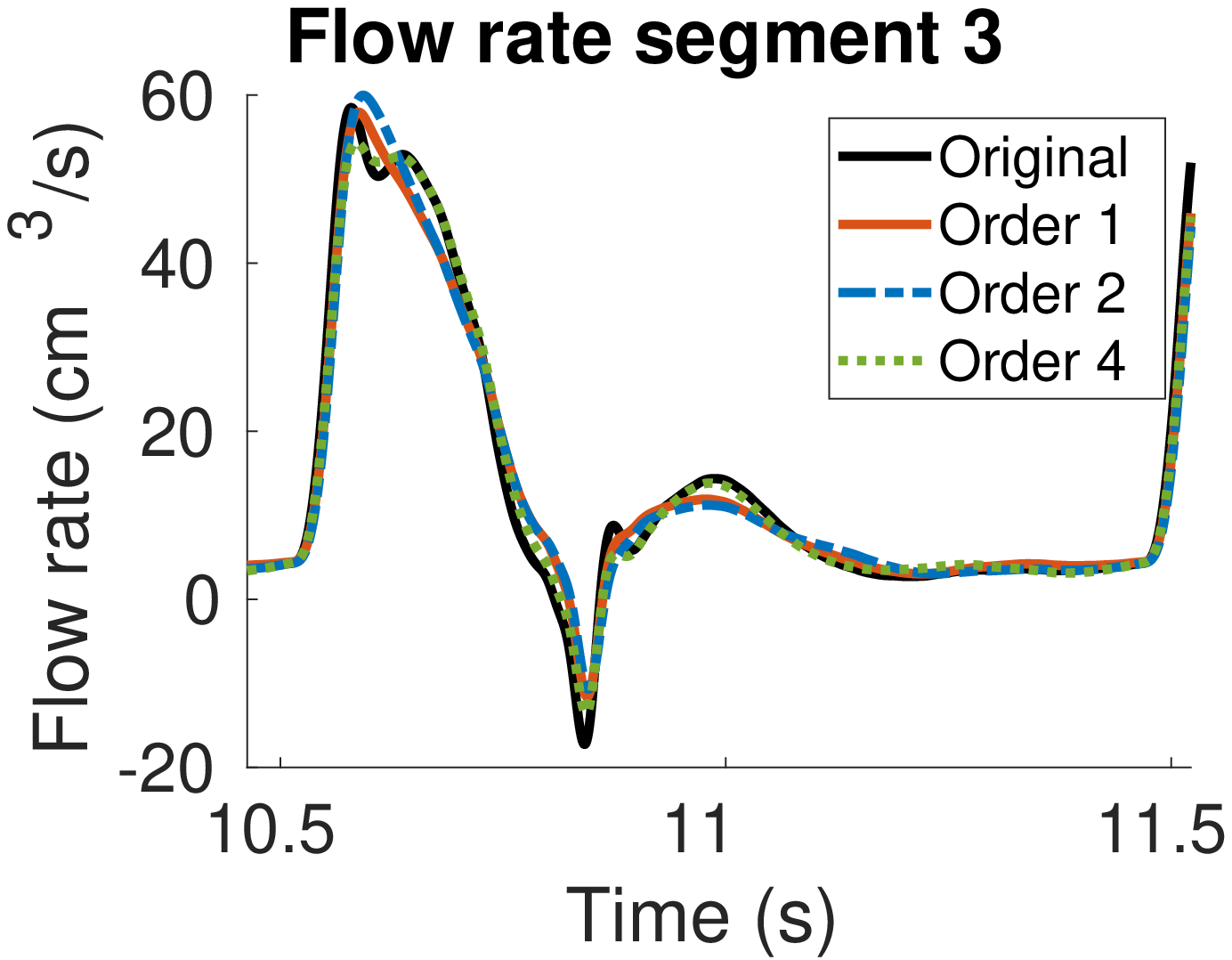}}
	\vspace*{0.2cm}
	\centerline{\includegraphics[width=0.49\columnwidth,clip]{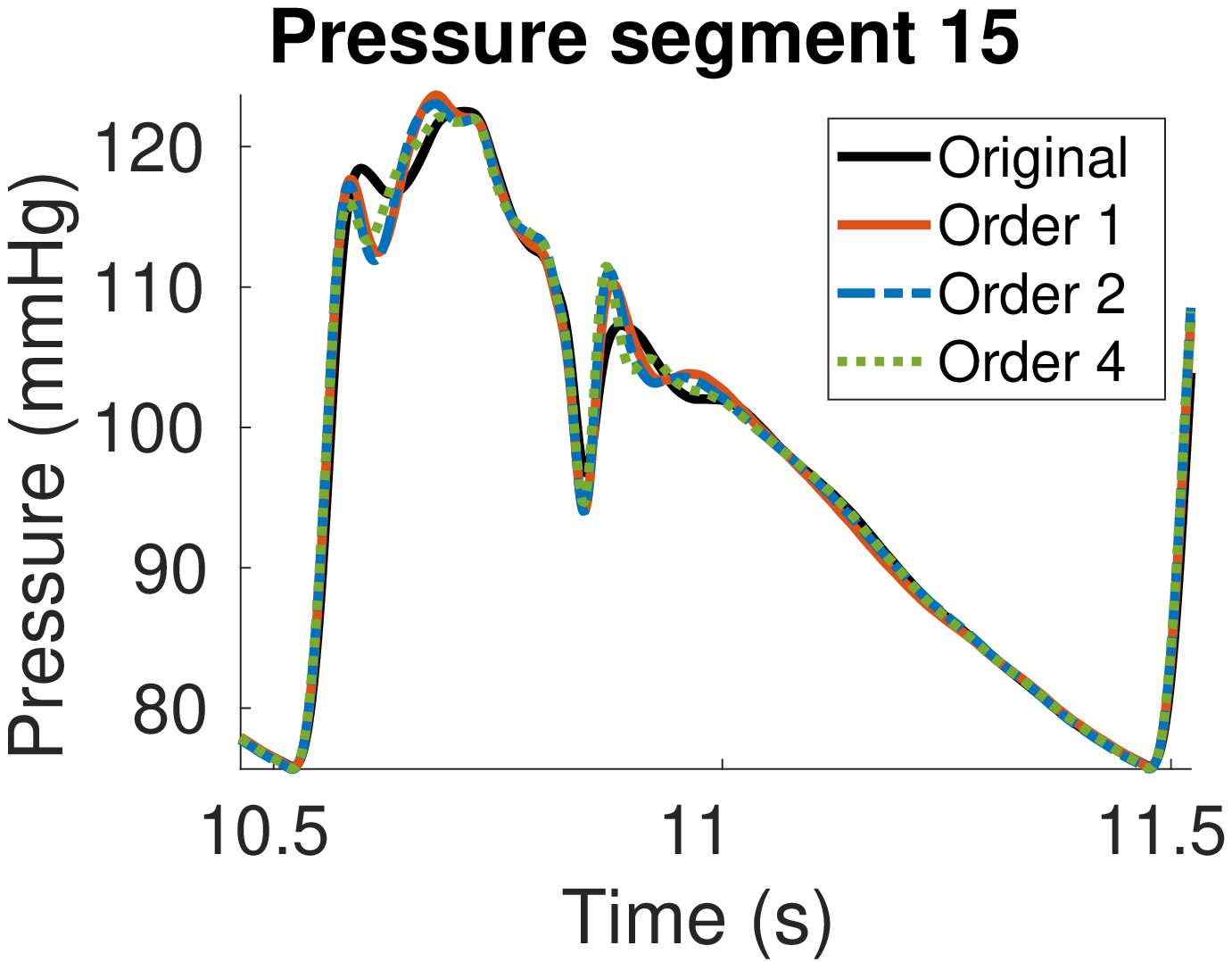}%
		\includegraphics[width=0.49\columnwidth,clip]{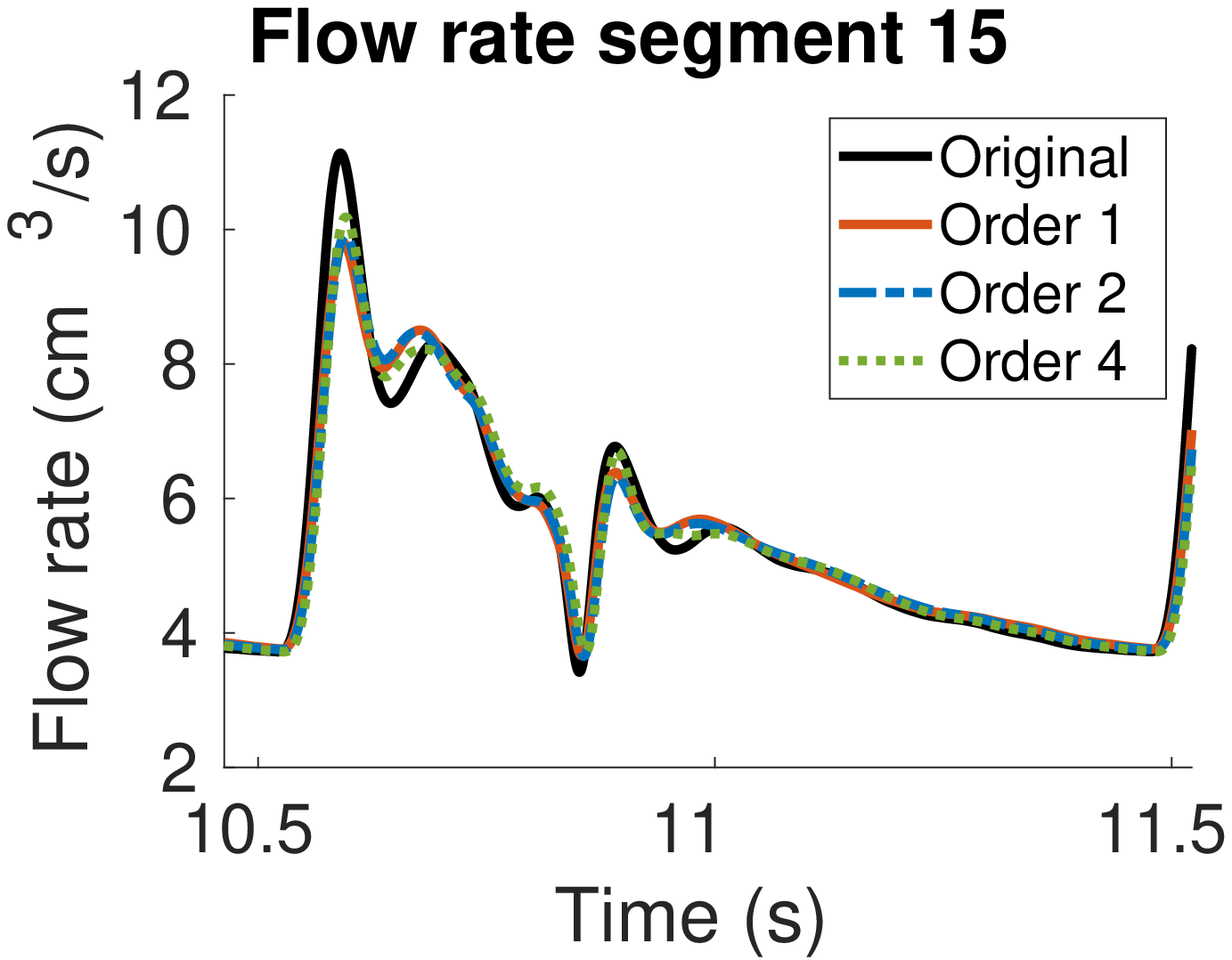}}%
	\vspace*{0.2cm}
	\centerline{\includegraphics[width=0.49\columnwidth,clip]{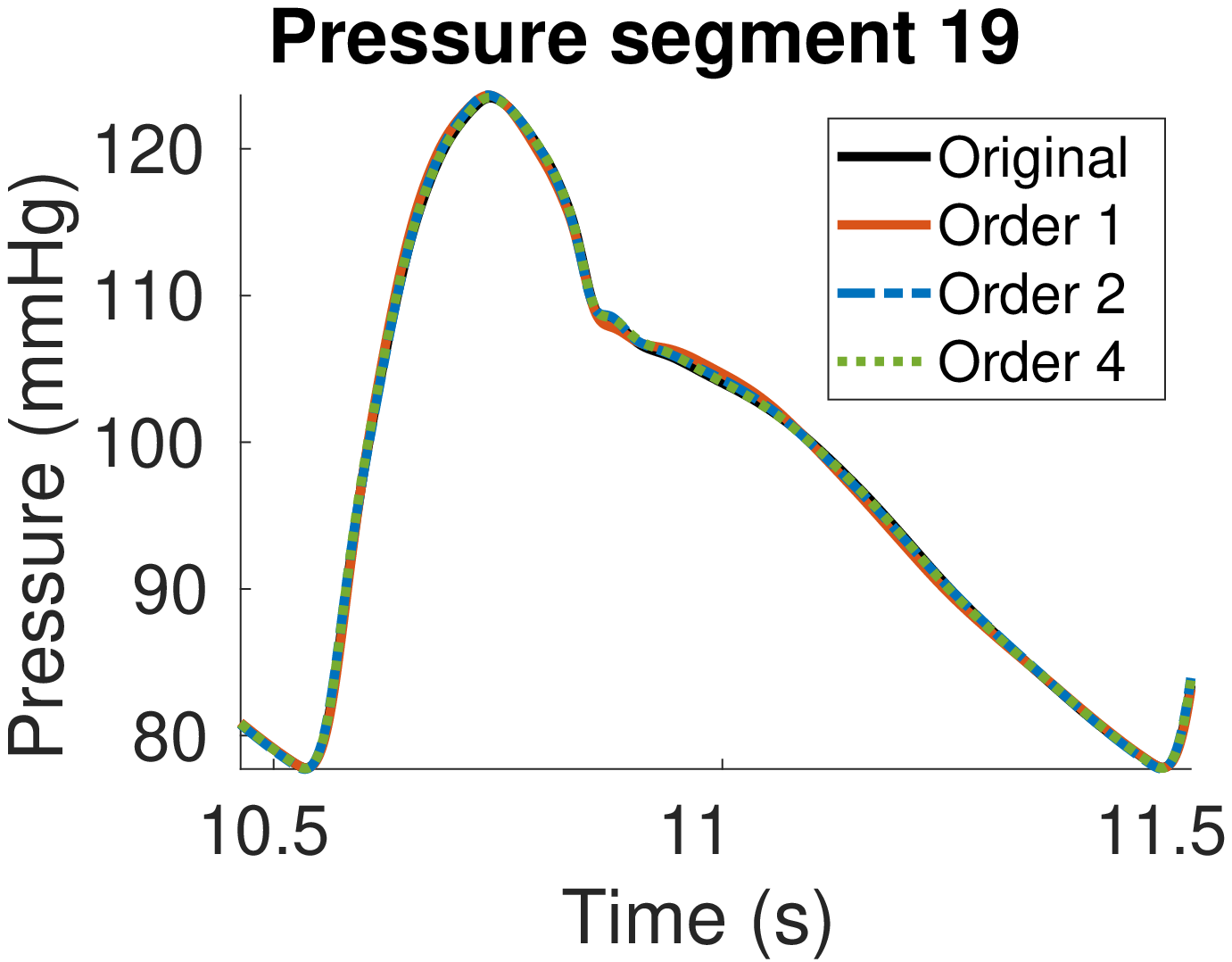}%
		\includegraphics[width=0.49\columnwidth,clip]{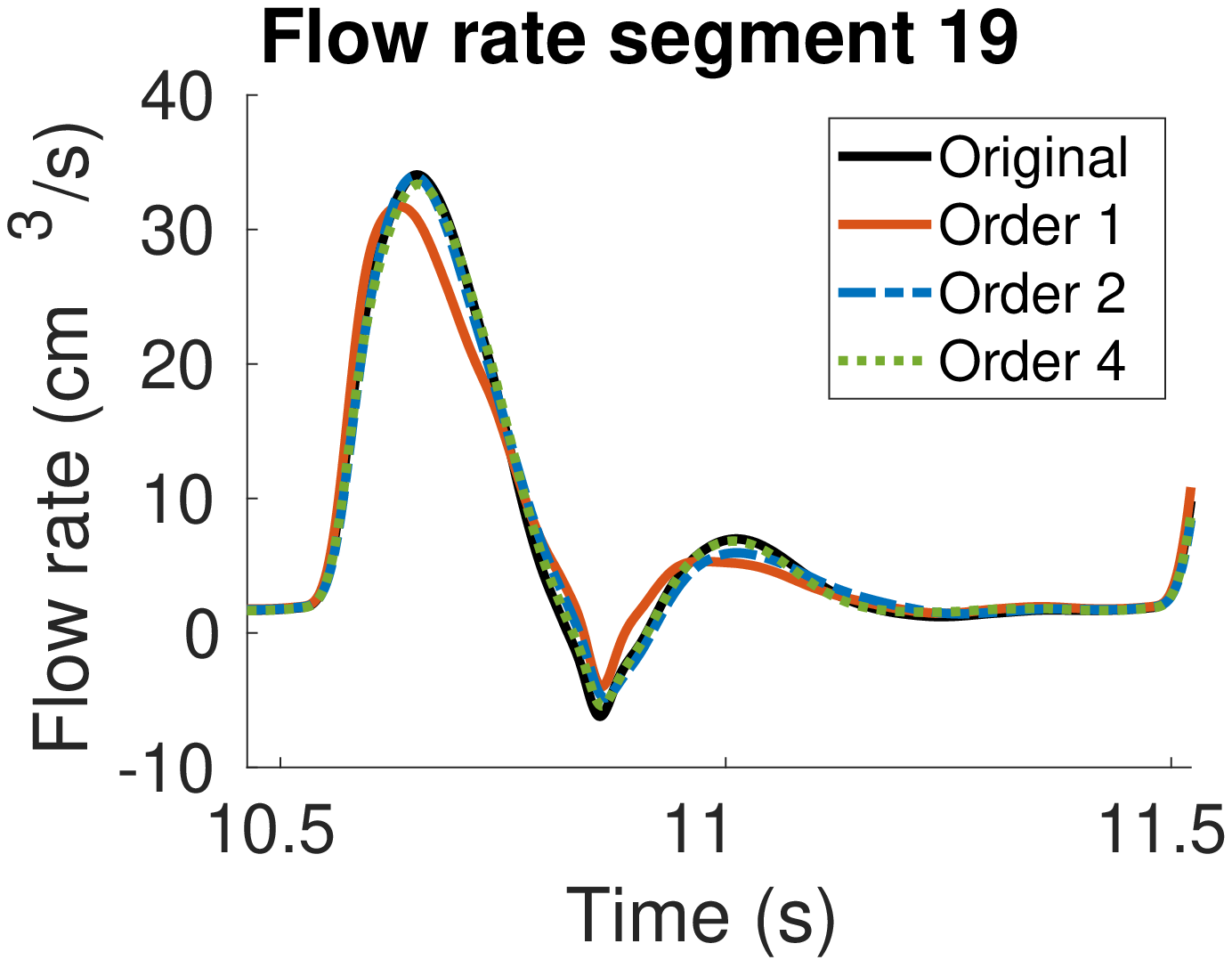}}%
	\vspace*{4cm}
	\end{figure}
	\begin{figure}\ContinuedFloat
	\centerline{\includegraphics[width=0.49\columnwidth,clip]{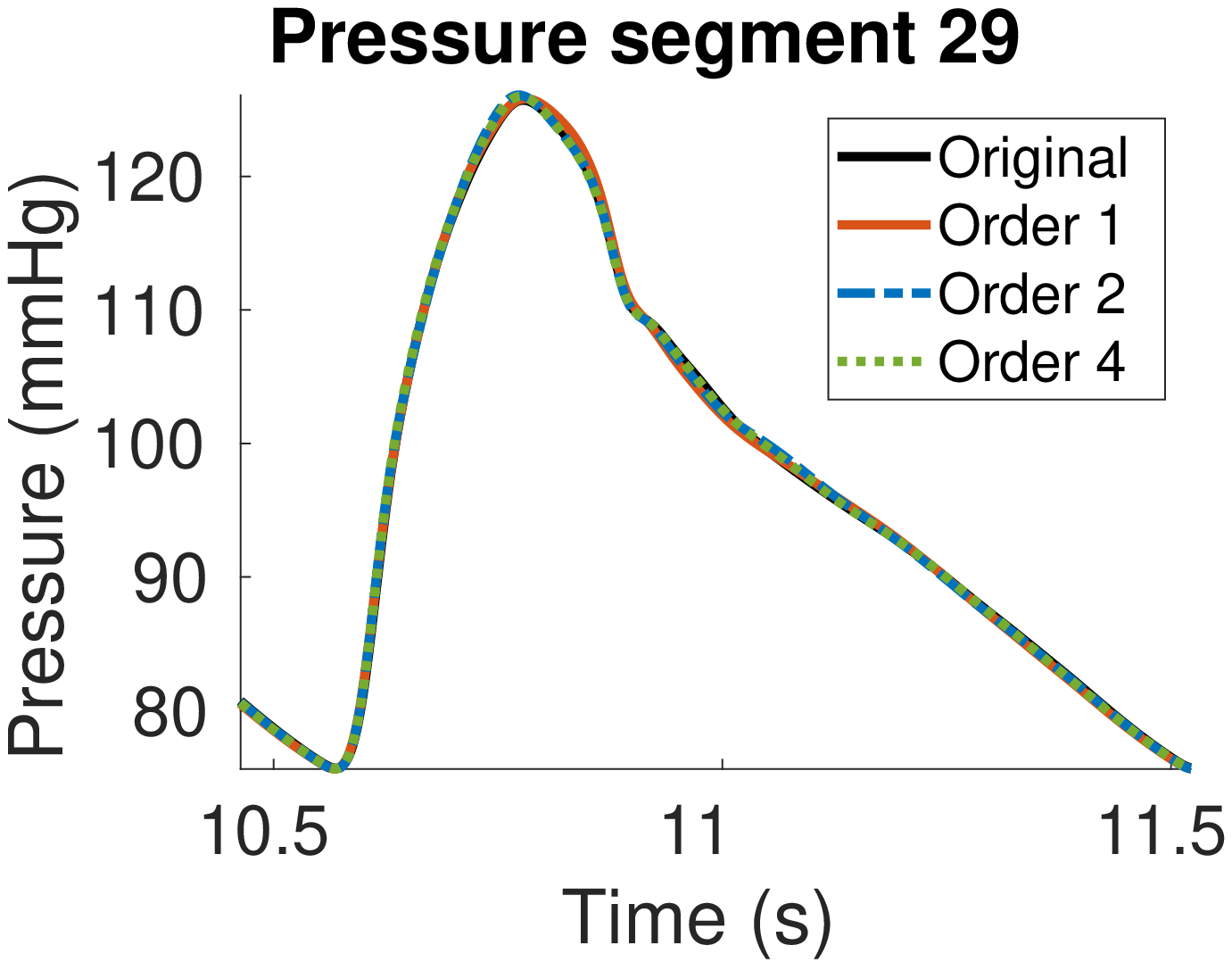}%
		\includegraphics[width=0.49\columnwidth,clip]{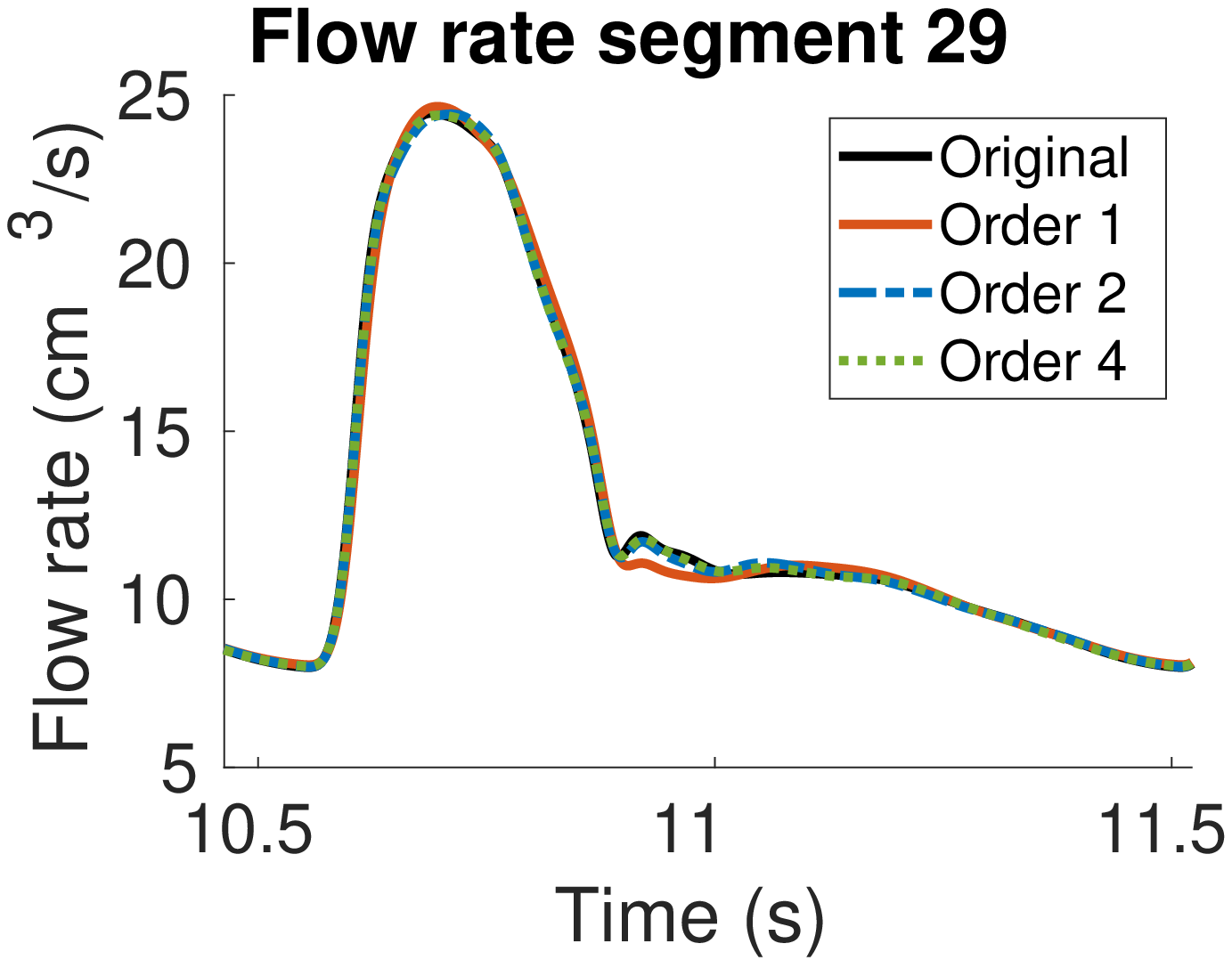}}%
	\vspace*{0.2cm}
	\centerline{\includegraphics[width=0.49\columnwidth,clip]{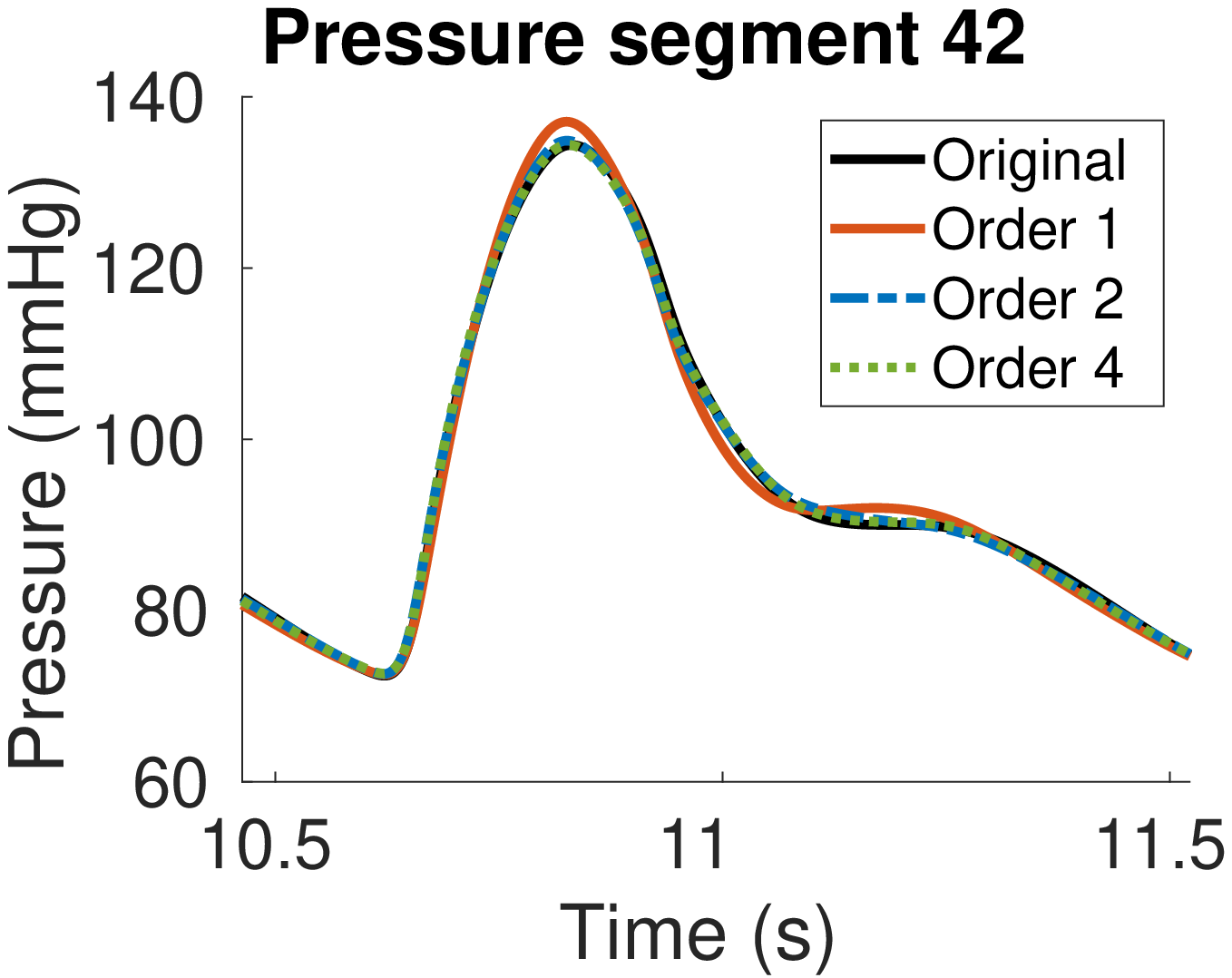}%
		\includegraphics[width=0.49\columnwidth,clip]{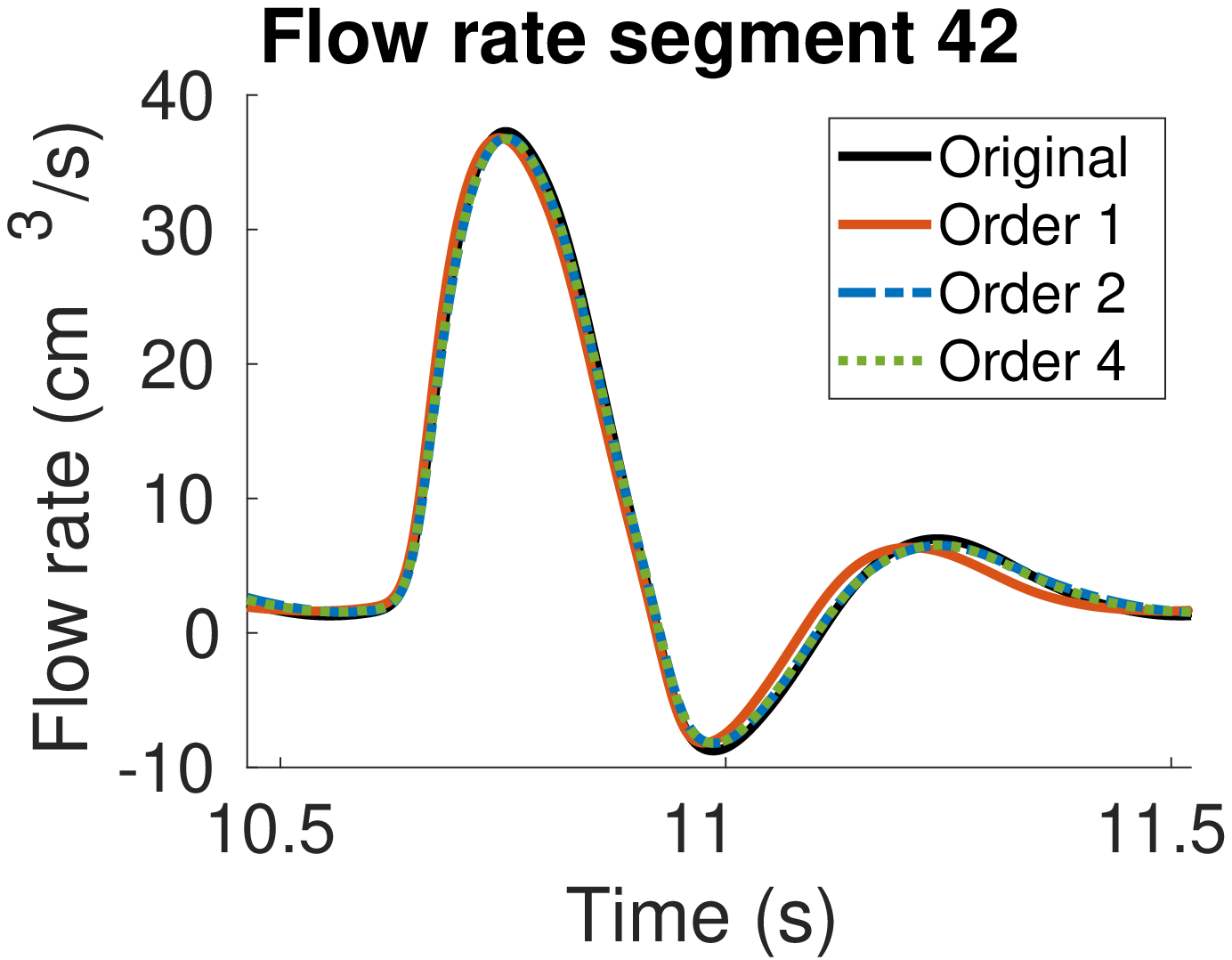}}%
	\caption{Comparison of pressure and flow waveforms in the 55-artery model (solid black curve) and in the 21-artery one with boundary conditions of different orders, estimated with Vector Fitting.} 
	\label{fig:accuracy_higher_order}
\end{figure}

\begin{table}
	\centering
	\caption{Average relative errors (\%) on pressure at truncation locations, with models of higher order (Sec.~\ref{subsec:higher})}
	\label{table:avg_errors_p_higher}
	\begin{tabular}{|c|ccccc|}
		\hline
		\multirow{2}{*}{Segment} & \multicolumn{5}{c|}{Average relative error (\%)}                                                                      \\ \cline{2-6} 
		& \multicolumn{1}{c|}{Order 1} & \multicolumn{1}{c|}{Order 2} & \multicolumn{1}{c|}{Order 4} & \multicolumn{1}{c|}{Order 6} & Order 8 \\ \hline
		3                        & \multicolumn{1}{c|}{0.30}  & \multicolumn{1}{c|}{0.12}  & \multicolumn{1}{c|}{0.078}  & \multicolumn{1}{c|}{0.084}  &  0.079 \\
		15                       & \multicolumn{1}{c|}{0.88}  & \multicolumn{1}{c|}{0.75}  & \multicolumn{1}{c|}{0.65}  & \multicolumn{1}{c|}{0.67}  &  0.67 \\
		19                       & \multicolumn{1}{c|}{0.30}  & \multicolumn{1}{c|}{0.1}  & \multicolumn{1}{c|}{0.069}  & \multicolumn{1}{c|}{0.079}  & 0.068  \\
		29                       & \multicolumn{1}{c|}{0.26}  & \multicolumn{1}{c|}{0.15}  & \multicolumn{1}{c|}{0.11}  & \multicolumn{1}{c|}{0.093}  & 0.079  \\
		42-43                       & \multicolumn{1}{c|}{1.1}  & \multicolumn{1}{c|}{0.36}  & \multicolumn{1}{c|}{0.31}  & \multicolumn{1}{c|}{0.23}  & 0.21  \\ \hline
	\end{tabular}
\end{table}

\begin{table}
	\centering
	\caption{Maximum relative errors (\%) on pressure at truncation locations, with models of higher order (Sec.~\ref{subsec:higher})}
	\label{table:max_errors_p_higher}
	\begin{tabular}{|c|ccccc|}
		\hline
		\multirow{2}{*}{Segment} & \multicolumn{5}{c|}{Maximum relative error (\%)}                                                                      \\ \cline{2-6} 
		& \multicolumn{1}{c|}{Order 1} & \multicolumn{1}{c|}{Order 2} & \multicolumn{1}{c|}{Order 4} & \multicolumn{1}{c|}{Order 6} & Order 8 \\ \hline
		3                        & \multicolumn{1}{c|}{0.67}  & \multicolumn{1}{c|}{0.59}  & \multicolumn{1}{c|}{0.46}  & \multicolumn{1}{c|}{0.53}  &  0.51 \\
		15                       & \multicolumn{1}{c|}{4.30}  & \multicolumn{1}{c|}{4.79}  & \multicolumn{1}{c|}{5.16}  & \multicolumn{1}{c|}{5.01}  &  4.80 \\
		19                       & \multicolumn{1}{c|}{0.62}  & \multicolumn{1}{c|}{0.44}  & \multicolumn{1}{c|}{0.33}  & \multicolumn{1}{c|}{0.34}  & 0.35  \\
		29                       & \multicolumn{1}{c|}{0.76}  & \multicolumn{1}{c|}{0.61}  & \multicolumn{1}{c|}{0.44}  & \multicolumn{1}{c|}{0.41}  & 0.42  \\
		42-43                       & \multicolumn{1}{c|}{2.81}  & \multicolumn{1}{c|}{1.33}  & \multicolumn{1}{c|}{1.17}  & \multicolumn{1}{c|}{1.07}  & 0.94  \\ \hline
	\end{tabular}
\end{table}

\begin{table}
	\centering
	\caption{Average relative errors (\%) on flow rate at truncation locations, with models of higher order (Sec.~\ref{subsec:higher})}
	\label{table:avg_errors_q_higher}
	\begin{tabular}{|c|ccccc|}
		\hline
		\multirow{2}{*}{Segment} & \multicolumn{5}{c|}{Average relative error (\%)}                                                                      \\ \cline{2-6} 
		& \multicolumn{1}{c|}{Order 1} & \multicolumn{1}{c|}{Order 2} & \multicolumn{1}{c|}{Order 4} & \multicolumn{1}{c|}{Order 6} & Order 8 \\ \hline
		3                        & \multicolumn{1}{c|}{3.14}  & \multicolumn{1}{c|}{3.63}  & \multicolumn{1}{c|}{1.80}  & \multicolumn{1}{c|}{1.70}  & 1.60  \\
		15                       & \multicolumn{1}{c|}{1.90}  & \multicolumn{1}{c|}{2.1}  & \multicolumn{1}{c|}{2.0}  & \multicolumn{1}{c|}{1.94}  &  1.90 \\
		19                       & \multicolumn{1}{c|}{3.41}  & \multicolumn{1}{c|}{1.67}  & \multicolumn{1}{c|}{1.0}  & \multicolumn{1}{c|}{0.95}  &  0.84 \\
		29                       & \multicolumn{1}{c|}{0.8}  & \multicolumn{1}{c|}{0.37}  & \multicolumn{1}{c|}{0.21}  & \multicolumn{1}{c|}{0.19}  & 0.17  \\
		42-43                       & \multicolumn{1}{c|}{2.5}  & \multicolumn{1}{c|}{1.07}  & \multicolumn{1}{c|}{0.97}  & \multicolumn{1}{c|}{0.61}  & 0.58  \\ \hline
	\end{tabular}
\end{table}

\begin{table}
	\centering
	\caption{Maximum relative errors (\%) on flow rate at truncation locations, with models of higher order (Sec.~\ref{subsec:higher})}
	\label{table:max_errors_q_higher}
	\begin{tabular}{|c|ccccc|}
		\hline
		\multirow{2}{*}{Segment} & \multicolumn{5}{c|}{Maximum relative error (\%)}                                                                      \\ \cline{2-6} 
		& \multicolumn{1}{c|}{Order 1} & \multicolumn{1}{c|}{Order 2} & \multicolumn{1}{c|}{Order 4} & \multicolumn{1}{c|}{Order 6} & Order 8 \\ \hline
		3                        & \multicolumn{1}{c|}{12.05}  & \multicolumn{1}{c|}{14.67}  & \multicolumn{1}{c|}{10.96}  & \multicolumn{1}{c|}{11.83}  & 10.80  \\
		15                       & \multicolumn{1}{c|}{13.71}  & \multicolumn{1}{c|}{15.61}  & \multicolumn{1}{c|}{16.47}  & \multicolumn{1}{c|}{16.12}  &  15.61 \\
		19                       & \multicolumn{1}{c|}{15.03}  & \multicolumn{1}{c|}{6.53}  & \multicolumn{1}{c|}{4.29}  & \multicolumn{1}{c|}{3.46}  &  3.58 \\
		29                       & \multicolumn{1}{c|}{4.45}  & \multicolumn{1}{c|}{1.22}  & \multicolumn{1}{c|}{0.92}  & \multicolumn{1}{c|}{0.98}  & 0.88  \\
		42-43                       & \multicolumn{1}{c|}{6.06}  & \multicolumn{1}{c|}{2.22}  & \multicolumn{1}{c|}{1.89}  & \multicolumn{1}{c|}{1.54}  & 1.40  \\ \hline
	\end{tabular}
\end{table}

\newpage
\section{Discussion} \label{sec:discussion}
Results presented in Section~\ref{sec:results}  show that Time-Domain Vector Fitting is able to estimate accurate lumped boundary conditions, making it a promising tool for cardiovascular modeling.
When employed to estimate Windkessel boundary conditions, Vector Fitting was able to accurately determine optimal values for the Windkessel parameters. The obtained results were further validated in presence of noisy measurements, where the proposed method provided accurate parameters starting from data with up to 20 dB of SNR, and under physiological changes of pressure and flow rates induced by changing levels of mental stress.
The results obtained with Vector Fitting were comparable to those attained with two other estimation methods presented in the literature.
The advantage of the proposed approach, however, is not only its accuracy, but also its ability to estimate an increasing number of parameters simultaneously and automatically. The proposed model parametrization based on the use of transfer functions, in fact, can be used to describe any linear dynamical system, allowing to generalize the model to differential relations of arbitrary order. The alternative solutions for higher order BCs proposed in the literature, instead, resort to specific circuit topologies, from which a generalization is difficult to obtain.
Thanks to the aforementioned properties of the TDVF method, it was possible to formulate and estimate in a systematic way boundary conditions of increasing order. 
For the case under analysis, consisting of a 55-artery model reduced to 21 arterial segments, boundary conditions with order up to 8 were estimated and compared, in order to assess the effect that the order of the boundary condition has on its ability to accurately approximate the downstream vasculature. Results showed that an order of 2 provides a significant increase in accuracy with respect to BCs of order 1, the most common choice up to now in the form of Windkessel models.
Orders above 4, instead, provided negligible improvements in terms of accuracy in the model of the systemic arterial system considered.

Higher order models could still be beneficial, for example, in the case of coronary circulation, for which the three-element Windkessel model has been deemed unable to reproduce the unique coronary haemodynamics, which is associated to non-negligible pressure pulsations at the myocardial level~\cite{frasch1996two}. Extended versions of the standard Windkessel model to better reproduce coronary haemodynamics have been proposed~\cite{jonavsova2021relevance}, but their use is still limited by the large number of parameters to be identified. In this sense, the Vector Fitting approach could remove this limitation, by automatically providing the parameters associated to a boundary condition of the desired complexity. This will be the subject of a future investigation.

\subsection{Limitations}
The estimation of boundary conditions with the proposed method requires time samples of both pressure and flow rate at the truncation location. This could be a limitation when pressure and flow rate measurements are not available simultaneously at the same location.
However, pressure and flow rate data are available when moving from a large model to one including a smaller portion of the cardiovascular system, like in the application considered in this paper. When moving to patient-specific models, both pressure and flow rate measurements can be obtained by means of \textit{in vivo} procedures and imaging techniques. In case pressure data are not available, pressure waveform generators can be used, which, given some patient-specific parameters (like brachial diastolic and systolic pressure) can generate realistic pressure waveforms~\cite{mariscal2021estimating, weber2011validation}.

In this work, the Vector Fitting method has been tested only on 1D models of the cardiovascular system. The possibility to extend the proposed approach to three-dimensional models will be investigated in future works.

\section{Summary and Conclusion} \label{sec:conclusion}
In this work, we proposed a new automated method based on the Time-Domain Vector Fitting algorithm for the estimation of boundary conditions for cardiovascular models. Starting from pressure and flow rate samples, this method can estimate boundary conditions corresponding to differential equations of increasing order.
First, the TDVF algorithm was used to automatically estimate 3WK boundary conditions, starting from a 1D model comprising the 55 main arteries of the human arterial system. The robustness of the estimation procedure was verified in presence of noisy data, with down to 20 dB of signal-to-noise ratio, and in presence of physiological changes of pressure and flow rate induced by high levels of mental stress.
Second, we proposed a generalization of the three-element Windkessel model to obtain boundary conditions of arbitrary order. We estimated higher order boundary conditions with TDVF, and we investigated the improvement in accuracy they provide with respect to the 3WK model. On the 55-artery model, experimental results showed that boundary conditions up to order 4 are able to model the downstream pressure and flow rate more accurately than the Windkessel model, while orders above 4 provided negligible improvements in term of accuracy.

\bibliographystyle{plain}
\bibliography{bibliography}

\end{document}